\documentclass[11pt]{article}
\usepackage{amsmath}

\usepackage{booktabs}

\usepackage{amssymb}

\usepackage{twemojis}

\usepackage[preprint]{acl}

\usepackage{times}
\usepackage{latexsym}

\usepackage[T1]{fontenc}

\usepackage{mathtools}

\usepackage[utf8]{inputenc}
\usepackage{microtype}

\usepackage{inconsolata}
\usepackage{amsthm}  

\newtheorem{proposition}{Proposition}

\usepackage[table]{xcolor}
\definecolor{SectionBg}{gray}{0.95} 
\definecolor{ReasoningBg}{HTML}{E8F0FE} 
\definecolor{OurMethodBg}{HTML}{FFF4E5}

\title{{\scalebox{1.6}{\twemoji{ear}}}Do Models Hear Like Us? \raisebox{-0.12\height}
Probing the Representational Alignment of Audio LLMs and Naturalistic EEG}

\author{
  \textbf{Haoyun Yang\textsuperscript{1,\ensuremath{*}}},
  \textbf{Xin Xiao\textsuperscript{1,\ensuremath{*}}},
  \textbf{Jiang Zhong\textsuperscript{1,\ensuremath{\dagger}}},
  \textbf{Yu Tian\textsuperscript{2}}
\\
  \textbf{DongXiaohua\textsuperscript{3}},
  \textbf{Yu Mao\textsuperscript{4}},
  \textbf{Hao Wu\textsuperscript{5}},
  \textbf{Kaiwen Wei\textsuperscript{1,\ensuremath{\dagger}}}
\\[2pt]
  \textsuperscript{1}School of Computer Science, Chongqing University, Chongqing, China \\
  \textsuperscript{2}Dept. of Comp. Sci. and Tech., Institute for AI, Tsinghua University, Beijing, China \\
  \textsuperscript{3}School of Economics and Business Administration, Chongqing University, Chongqing, China \\
  \textsuperscript{4}School of Artificial Intelligence, Southwest University, Chongqing, China \\
  \textsuperscript{5}The First Affiliated Hospital of Chongqing Medical University, Chongqing, China \\  
  \texttt{\{20241536,20241401023\}@}
  {\begingroup\hypersetup{urlcolor=black}
   \href{mailto:@stu.cqu.edu.cn}{\texttt{stu.cqu.edu.cn}}
   \endgroup}\;
  \texttt{\{zhongjiang,weikaiwen\}@}
  {\begingroup\hypersetup{urlcolor=black}
   \href{mailto:@cqu.edu.cn}{\texttt{cqu.edu.cn}}
   \endgroup}
}

\begin{document}
\maketitle

\newcounter{titlefootnote}
\setcounter{titlefootnote}{\value{footnote}}
\begingroup
\renewcommand{\thefootnote}{\fnsymbol{footnote}}
\setcounter{footnote}{0}
\footnotetext[1]{Equal contributions.}
\footnotetext[2]{Corresponding authors.}
\endgroup
\setcounter{footnote}{\value{titlefootnote}}

\begin{abstract}
Audio Large Language Models (Audio LLMs) have demonstrated strong capabilities in integrating speech perception with language understanding. However, whether their internal representations align with human neural dynamics during naturalistic listening remains largely unexplored. In this work, we systematically examine layer-wise representational alignment between 12 open-source Audio LLMs and Electroencephalogram (EEG) signals across 2 datasets. Specifically, we employ 8 similarity metrics, such as Spearman-based Representational Similarity Analysis (RSA), to characterize within-sentence representational geometry. Our analysis reveals 3 key findings: (1) we observe a rank-dependence split, in which model rankings vary substantially across different similarity metrics; (2) we identify spatio-temporal alignment patterns characterized by depth-dependent alignment peaks and a pronounced increase in RSA within the 250-500 ms time window, consistent with N400-related neural dynamics; (3) we find an affective dissociation whereby negative prosody, identified using a proposed Tri-modal Neighborhood Consistency (TNC) criterion, reduces geometric similarity while enhancing covariance-based dependence. These findings provide new neurobiological insights into the representational mechanisms of Audio LLMs.

\end{abstract}

\section{Introduction}
Human speech comprehension hierarchically transforms acoustic signals into stable linguistic and semantic representations across multiple timescales \citep{hickok2007cortical_org, giraud2012oscillations_speech,ding2016hierarchical_tracking}. A central goal of brain–model alignment is to determine whether and when computational speech-language representations reflect neural dynamics during naturalistic listening, providing a principled benchmark for comparing audio–language models and informing human-compatible and brain-aware speech systems \citep{defossez2023decoding_speech}.

\begin{figure}[t]
\centering
\includegraphics[width=0.7\columnwidth]{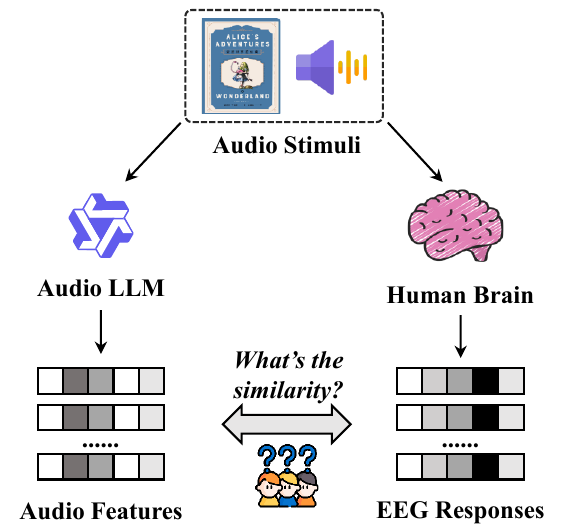}
\caption{The motivation of this paper: exploring the similarity of human brain EEG signals and Audio LLMs representations through various similarity metrics.}
\label{fig:intro}
\end{figure}

Recently, model–brain alignment is increasingly evaluated via layer-wise Representational Similarity Analysis (RSA) or encoding models \citep{sartzetaki2024hundred, gao2025llm_alignment}. While speech research has predominantly used Temporal Response Functions (TRF) to model acoustic coupling, the representational-space alignment of instruction-following Audio Large Language Models (Audio LLMs) has only begun to receive attention \citep{wu2025distinct_audiolalm}. This aspect is important because naturalistic speech intertwines linguistic and prosodic information, which modulates semantic integration within the canonical N400 window \citep{kutas2011n400review}. Moreover, alignment results vary considerably across similarity metrics \citep{soni2024similarity_impact}, highlighting the need for a robust, multi-metric evaluation to fully capture these complex neural dynamics.

However, the majority of existing model–brain alignment studies have focused on vision and text domains using functional Magnetic Resonance Imaging (fMRI), while systematic investigations of speech processing with Electroencephalography (EEG) remain comparatively scarce. Unlike fMRI, EEG offers superior temporal resolution, enabling the capture of rapidly unfolding neural dynamics~\citep{burle2015eeg_resolution}. Motivated by these considerations, we examine whether modern Audio LLMs develop intermediate representations whose geometry aligns with EEG during naturalistic listening, and characterize how such alignment varies across model depth, time, and prosodic conditions.

To evaluate the alignment between Audio LLMs and human EEG responses, we temporally align model states with sentence-level EEG segments and construct within-sentence representational dissimilarity matrices \citep{kriegeskorte2008rsa}. We quantify the similarity between EEG and Audio LLM using a comprehensive suite of 8 metrics, such as Pearson RSA~\cite{kriegeskorte2008rsa}, linear and kernel Centered Kernel Alignment (CKA) \citep{kornblith2019cka}. To further investigate condition-specific effects, we stratify sentences by affective prosody eGeMAPS features~\citep{eyben2016egemaps}. In addition, we introduce Tri-modal Neighborhood Consistency (TNC), a metric that requires EEG–model alignment to be jointly supported by the underlying acoustic stimulus in both affective and prosodic dimensions.

We conduct experiments with 12 Audio LLMs on 2 EEG datasets, and observe consistent patterns: (1) alignment is strongly depth-dependent, but the layer of maximal alignment varies by metric; (2) rank-based geometric measures and dependence-based measures peak at different depths, yielding a robust rank–dependence split in cross-model rankings; (3) time-resolved scalp analyses localize the strongest RSA to the 250–500 ms window, consistent with canonical N400-range semantic integration latencies \citep{kutas1980senseless_sentences,kutas2011n400review}; and (4) prosody-aware analyses reveal an affective dissociation in which negative prosody weakens rank-based geometric agreement while strengthening covariance-style dependence, suggesting that affect reshapes temporal neighborhood structure rather than uniformly scaling coupling strength. In summary, the main contributions of this work are as follows:

(1) We perform a multi-metric, layer-wise EEG–Audio LLM similarity study across 2 datasets using 8 complementary similarity measures. To the best of our knowledge, this is the first work to systematically examine representational similarity between Audio LLMs and EEG signals.

(2) The time-resolved analysis reveals depth- and time-dependent alignment patterns, where rank-based geometry and dependence-based coupling capture complementary alignment signals and peak at different layers while converging in language-relevant temporal windows.

(3) We introduce a prosody-aware tri-modal analysis with a Tri-modal Neighborhood Consistency (TNC) criterion. Experiments reveal a robust affective dissociation in which negative prosody systematically modulates EEG–model alignment.

\section{Related Work}
\subsection{Neural Basis of Speech Comprehension}
Spoken language comprehension unfolds hierarchically across multiple timescales, mapping acoustic signals to phonological and higher-level linguistic representations \citep{poeppel2003ast,giraud2012oscillations_speech,ding2016hierarchical_tracking}. This process engages a left-lateralized frontotemporal network linking posterior superior temporal and inferior frontal regions via pathways such as the arcuate fasciculus \citep{schroen2023_temporal_interplay,friederici2026_structural_networks}. 
EEG studies indicate that speech comprehension reflects not only acoustic tracking but also semantic processing, with N400-range activity serving as a canonical marker of lexical–semantic integration during spoken language understanding \citep{kutas1980senseless_sentences,holcomb1990_semantic_priming,grisoni2021_spp_n400}. As prosody can modulate N400 dynamics, we examine EEG–model alignment in the 250-500 ms window and adopt prosody-aware analyses to disentangle linguistic alignment from acoustics- or prosody-driven effects \citep{schirmer2002_sex_emotional_prosody,kotz2007_prosody_semantics,themistocleous2025_prosody_ale}.

\begin{figure*}[t!]
    \centering
    \includegraphics[width=0.9\linewidth]{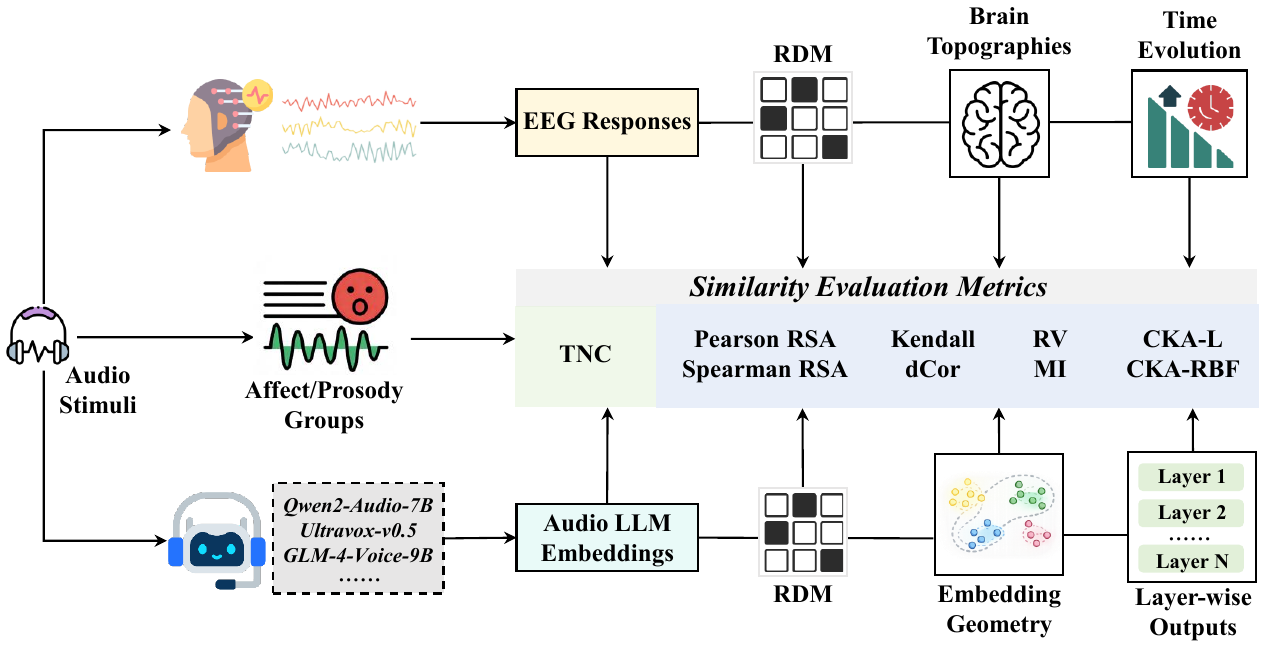}
    \caption{The workflow for EEG–Audio LLM similarity analysis. EEG and Audio LLM representations are preprocessed and temporally aligned, compared using 8 complementary similarity measures, and further analyzed with a tri-modal neighborhood consistency (TNC) metric for affective and prosodic alignment.}
    \label{fig:framework}
\end{figure*}

\subsection{Brain Similarity with LLMs}
Speech–brain alignment has largely relied on scalar information-theoretic predictors, such as word surprisal, and encoding or TRF-style models to explain EEG responses and N400-like effects during naturalistic comprehension \citep{frank2015erp_information,brennan2019hierarchical_predictions,weissbart2020tracking_surprisal,broderick2021ageing_strategy,crosse2016mtrf_toolbox,broderick2018semantic}. In fMRI, voxel-wise modeling of natural speech has revealed cortical hierarchies spanning spectral to semantic representations \citep{huth2016natural_speech,deheer2017hierarchical_cortical_speech}.
By contrast, text and vision studies increasingly use layer-wise representational alignment methods, such as RSA and encoding models, to probe depth-dependent brain correspondence \citep{kriegeskorte2008rsa,jain2018context_fmri,schrimpf2021neural_arch}. Although similar analyses exist for speech models, they primarily target ASR or self-supervised encoders rather than instruction-following Audio LLMs \citep{baevski2020wav2vec2,hsu2021hubert,vaidya2022selfsupervised_audio_cortex}. Consequently, systematic EEG–Audio LLM similarity analyses remain scarce, and we address this gap with layer-resolved, multi-electrode comparisons across modern Audio LLMs.

\section{Method}
\label{sec:method}

We examine how representational similarity evolves across layers of Audio LLMs and how these layer-wise geometries correspond to EEG signals during naturalistic listening. The workflow is illustrated in Figure~\ref{fig:framework}: First, we preprocess EEG signals to derive complementary neural feature views. Second, we extract layer-wise hidden states from multiple Audio LLMs and temporally align them with EEG recordings. Third, we quantify cross-modal similarity using 8 complementary similarity metrics and assess statistical significance via permutation testing. Finally, we introduce a tri-modal consistency index (TNC) metric and analyze how similarity varies across speech conditions, such as emotional and prosodic partitions.

\subsection{Data Preprocessing}
\paragraph{Sentence-level EEG Segmentation.}
For each sentence audio stimulus $s$, we extract the EEG segment spanning sentence onset to offset.
Within each segment, we apply $z$-score normalization independently for each electrode to standardize channel scales \citep{hastie2009esl,murphy2012pml}.
The resulting sentence-locked EEG is represented as
$\mathbf{U}^{\mathrm{eeg,raw}}_s \in \mathbb{R}^{N_s \times C}$,
where $N_s$ denotes the number of sampled time points and $C$ the number of retained scalp electrodes, with $c \in \{1,\ldots,C\}$ indexing individual electrodes.

\paragraph{Audio LLM Feature Extraction.}
For each sentence $s$, we feed its audio waveform into a pretrained Audio LLM and extract hidden representations from all layers. Most Audio LLMs use a transformer architecture \citep{vaswani2017attention}, where an audio encoder converts the waveform into a sequence of audio tokens that are projected and processed by $L$ stacked transformer blocks in the backbone. 
We index model depth by $l \in \{0,\ldots,L-1\}$, where $l=0$ corresponds to the output of the first Transformer decoder layer and larger $l$ denote progressively deeper layers.
Let $T_a$ denote the number of audio tokens for sentence $s$, and $D$ the hidden dimensionality.
The token-synchronous representation at layer $l$ of LLM is defined as:
\begin{equation}
\mathbf{V}^{\mathrm{llm},l}_s
=
\bigl[\mathbf{v}^{\mathrm{llm},l}_{s,1}, \ldots, \mathbf{v}^{\mathrm{llm},l}_{s,T_a}\bigr]^\top
\in \mathbb{R}^{T_a \times D},
\end{equation}
where $\mathbf{v}^{\mathrm{llm},l}_{s,t} \in \mathbb{R}^{D}$ denotes the model state at token time $t$.
In addition, we further apply Principal Component Analysis (PCA) and retain the top $k$ components \citep{jolliffe2002pca}.

\paragraph{Temporal Alignment.}
The sentence-locked EEG epoch contains $N_s$ time points, whereas the corresponding Audio LLM representation consists of $T_a$ token-level states.
To enable token-synchronous comparison, we resample each electrode time series onto the model token grid of length $T_a$ using electrode-wise linear interpolation \citep{oppenheim2009dtsp,smith2010pasp}.
This yields the aligned EEG representation $\mathbf{U}^{\mathrm{eeg}}_s \in \mathbb{R}^{T_a \times C}$.

\subsection{EEG--LLM Similarity Measures}
\label{sec:similarity}

For each sentence, we compare time-aligned EEG sequences with representations from an Audio LLM layer using 8 similarity measures: Pearson RSA~\cite{kriegeskorte2008rsa}, Spearman RSA~\cite{spearman1904_association}, and Kendall $\tau_b$ \citep{kendall1945tiesties}, distance correlation (dCor)~\cite{szekely2007dcor}, RV coefficient~\cite{robert1976rv}, Gaussian mutual information proxy (MI)~\cite{cover2006it}, and centered kernel alignment with linear (CKA-L) and RBF kernels (CKA-RBF)~\cite{kornblith2019cka}. These metrics capture distinct aspects of correspondence, spanning linear and nonlinear dependence as well as rank-based and covariance-based structure. 

\paragraph{Representational Similarity Analysis (RSA).}
RSA evaluates whether features of LLM and EEG share a common relational geometry over time, i.e., whether pairs of time points that are similar or dissimilar in one modality exhibit consistent similarity relations in the other \citep{kriegeskorte2008rsa}.

Take the EEG feature as an example,
$\mathbf{U}^{\mathrm{eeg}}_s=[\mathbf{u}^{\mathrm{eeg}}_{s,1},\ldots,\mathbf{u}^{\mathrm{eeg}}_{s,T_a}]\in\mathbb{R}^{T_a\times C}$,
where $\mathbf{u}^{\mathrm{eeg}}_{s,i}\in\mathbb{R}^{C}$ is the EEG feature vector at token-aligned time step $i$, we implement RSA by constructing representational dissimilarity matrices (RDMs)~\cite{nili2014rsatoolbox} using correlation distance:
\begin{equation}
\mathrm{RDM}^{\mathrm{eeg}}_s(i,j)=1-\mathrm{corr}\!\left(\mathbf{u}^{\mathrm{eeg}}_{s,i},\mathbf{u}^{\mathrm{eeg}}_{s,j}\right),
\label{eq:rdm}
\end{equation}
where $i,j \in \{1,\ldots,T_a\}$ index token-aligned time steps, and $\mathrm{corr}(\cdot,\cdot)$ denotes Pearson correlation \citep{pearson1895_regression_inheritance} across feature dimensions.

Following~\citet{kriegeskorte2008rsa}, we vectorize the strictly upper-triangular entries of each RDM to obtain the EEG representation
$\mathbf{r}^{\mathrm{eeg}}_s=\operatorname{vec}_{\triangle}(\mathrm{RDM}^{\mathrm{eeg}}_s)$
and the Audio LLM representation
$\mathbf{r}^{\mathrm{llm},l}_s=\operatorname{vec}_{\triangle}(\mathrm{RDM}^{\mathrm{llm},l}_s)$,
where $\operatorname{vec}_{\triangle}(\cdot)$ stacks entries with $i<j$, excluding the diagonal and redundant symmetric terms.
The RSA score is defined as the correlation between these vectorized RDMs, using Spearman correlation to provide rank-based robustness, or Pearson correlation to quantify linear agreement \citep{spearman1904_association,pearson1895_regression_inheritance}:
\begin{equation}
\small
\begin{aligned}
\rho^{\mathrm{eeg,llm},l}_s &= \mathrm{Spearman}\!\left(\mathbf{r}^{\mathrm{eeg}}_s,\,\mathbf{r}^{\mathrm{llm},l}_s\right),\\
r^{\mathrm{eeg,llm},l}_s &= \mathrm{Pearson}\!\left(\mathbf{r}^{\mathrm{eeg}}_s,\,\mathbf{r}^{\mathrm{llm},l}_s\right).
\end{aligned}
\label{eq:rsa}
\end{equation}

\paragraph{Complementary Dependence Measures.}
RSA captures second-order geometry by comparing pairwise relations among time points. To complement this view, we report 6 additional dependence measures, each highlighting a different notion of cross-modal association: \emph{Kendall’s $\tau_b$} quantifies monotonic rank agreement while correcting for ties, providing robustness when repeated values occur. \emph{Distance correlation (dCor)} detects general statistical dependence, including nonlinear relationships. The \emph{RV coefficient} measures linear association between two multivariate representations in a covariance-based manner. A \emph{Gaussian mutual-information proxy} captures nonlinear dependence from an information-theoretic perspective. Finally, \emph{centered kernel alignment (CKA)} compares representation spaces via kernel similarity; we report both \emph{linear CKA} and \emph{RBF CKA} to distinguish linear from nonlinear alignment. Formal details are given in Appendix~\ref{app:metric-definitions}.

\paragraph{Permutation-based Significance Testing.}
A similarity score is meaningful only if it exceeds chance under the null hypothesis of no cross-modal temporal correspondence.
We therefore assess statistical significance using a nonparametric time-shuffle permutation test, which disrupts temporal alignment by permuting time indices in one modality \citep{nichols2002nonparametric}.
Implementation details, including the number of permutations and $p$-value estimation, are provided in Appendix~\ref{app:impl_details}.

\subsection{Affect and Prosodic Similarity Measures}
To further assess how LLM-EEG alignment varies across speech conditions (e.g., affective and prosodic partitions), we note that pairwise EEG--LLM similarity can be inflated by indirect or shared confounds. Therefore, we require consistent representational geometry across acoustics, EEG, and Audio LLM representations, and introduce a Tri-modal Neighborhood Consistency (TNC) measure.

Formally, we adopt a Spearman RSA-style framework based on RDMs \citep{kriegeskorte2008rsa,nili2014rsatoolbox}.
Using Eq.~\ref{eq:rdm} and Eq.~\ref{eq:rsa}, we compute the three pairwise Spearman RSA correlations among acoustics, EEG, and the layer-$l$ Audio LLM representations.
We then define TNC as:
\begin{equation}
\begin{aligned}
\mathrm{TNC}^{\,l}_s
&=\frac{1}{3}\Big(
\big[\rho^{\mathrm{ac,eeg}}_s\big]^2
+\big[\rho^{\mathrm{eeg,llm},l}_s\big]^2 \\
&\qquad\quad
+\big[\rho^{\mathrm{ac,llm},l}_s\big]^2
\Big),
\end{aligned}
\label{eq:tnc}
\end{equation}
where the superscripts indicate the modality pair used to compute the corresponding RSA correlation:
$\mathrm{ac,eeg}$ denotes acoustic--EEG similarity, $\mathrm{eeg,llm}$ denotes EEG--model-embedding similarity at layer $l$,
and $\mathrm{ac,llm}$ denotes acoustic--model-embedding similarity at layer $l$.

By jointly constraining alignment across stimulus acoustics, neural responses, and model representations, TNC provides a more conservative and robust estimate of cross-modal correspondence, mitigating the influence of modality-specific noise and indirect effects \citep{walther2016reliability,kriegeskorte2008rsa}. $\mathrm{TNC}^{\,l}_s$ is high only when \emph{all three} modality pairs exhibit mutually consistent within-sentence representational geometry. A detailed mathematical justification of these properties is provided in Appendix~\ref{app:tnc-theory}.

\paragraph{Sentence Partitioning by Affect and Prosody.}
To examine whether cross-modal alignment varies across speech conditions, we construct two complementary sentence groupings: an \emph{affect-based} partition derived from an acoustic valence proxy, and a \emph{prosody-based} partition obtained via clustering over multi-dimensional prosodic descriptors.

(1) \textit{Affect-based Groups.}
We compute sentence-level acoustic descriptors using openSMILE / eGeMAPS \citep{eyben2010opensmile,eyben2016egemaps} and apply $z$-score normalization across sentences \citep{murphy2012pml}.
Using standardized features, we form an acoustic valence proxy:
\begin{equation}
\begin{aligned}
\mathrm{valence}(s)= a_1\,z_{\mathrm{pitch}} + a_2\,z_{\alpha} + a_3\,z_{\mathrm{hammarberg}},
\end{aligned}
\end{equation}
where $\mathrm{pitch}$ corresponds to $F_0$, $\alpha$ denotes a selected eGeMAPS / openSMILE descriptor, and $\mathrm{hammarberg}$ is the Hammarberg index \citep{hammarberg1980abnormal,eyben2016egemaps}.

We then assign each sentence to one of three affect groups using a symmetric threshold $\tau_v$:
positive if $\mathrm{valence}(s)>\tau_v$, negative if $\mathrm{valence}(s)<-\tau_v$, and neutral otherwise.
Within these affect groups, we report the average EEG--LLM similarity results of the 8 measures.

(2) \textit{Prosody-based Groups.} To capture broader prosodic variation beyond low-dimensional affect proxies, we construct a prosodic feature vector $\mathbf{p}_s$ from duration, energy, pitch, voicing, ZCR, and spectral-shape descriptors (Appendix~\ref{app:partition-details}), $z$-score normalize across sentences, and cluster sentences with $K$-means \citep{macqueen1967kmeans}. For each cluster, we compute sentence-level TNC (Eq.~\ref{eq:tnc}) and summarize it by the cluster mean and standard deviation to compare tri-modal alignment across prosodic regimes. Specific values and details are provided in Appendix~\ref{app:impl_details}.

\section{Experiments}

\begin{table*}[t]
\centering
\small 
\scriptsize
\setlength{\tabcolsep}{3pt}
\renewcommand{\arraystretch}{1.15}

\resizebox{0.95\textwidth}{!}{
\begin{tabular}{lcccccccc}
\toprule
Model 
& Pearson RSA $\uparrow$ 
& Spearman RSA $\uparrow$ 
& Kendall $\uparrow$ 
& dCor $\uparrow$ 
& RV $\uparrow$ 
& MI $\uparrow$ 
& CKA-L $\uparrow$ 
& CKA-RBF $\uparrow$ \\
\midrule

\rowcolor{ReasoningBg}
\multicolumn{9}{c}{\textit{Alice in Wonderland Dataset}~\citep{brennan2019hierarchical_predictions}} \\
\midrule
Audio-Flamingo-3 & \textbf{0.2857 }& 0.1748 & 0.1092 & 0.3342 & 0.2233 & 0.0193 & 0.2233 & 0.3112 \\
Baichuan-Audio-Base & 0.2028 & 0.1527 & 0.1107 & 0.3774 & 0.2216 & 0.0190 & 0.2216 & 0.3256 \\
Baichuan-Audio-Instruct & 0.2286 & 0.1690 & 0.1125 & 0.4067 & 0.2394 & 0.0205 & 0.2394 & 0.3486 \\
GLM-4-Voice-9B & 0.2597 & \textbf{0.2031} & 0.1376 & 0.3008 & 0.2011 & 0.0162 & 0.2011 & 0.2783 \\
Granite-Speech-3.3-8B & 0.2531 & 0.1949 & \textbf{0.1387} & 0.3813 & 0.2226 & 0.0179 & 0.2226 & 0.3154 \\
Llama-3.1-8B-Omni & 0.2396 & 0.1765 & 0.1144 & 0.4580 & 0.2777 & \textbf{0.0268} & 0.2777 & 0.3960 \\
MiniCPM-o-2\_6 & 0.2376 & 0.1749 & 0.1229 & 0.3113 & 0.1999 & 0.0158 & 0.1999 & 0.2762 \\
Qwen2-Audio-7B & 0.2306 & 0.1848 & 0.1248 & 0.2962 & 0.1788 & 0.0139 & 0.1788 & 0.2540 \\
Qwen2-Audio-7B-Instruct & 0.2368 & 0.1773 & 0.1334 & 0.3138 & 0.1931 & 0.0151 & 0.1931 & 0.2767 \\
SpeechGPT-2.0-preview-7B & 0.2527 & 0.1931 & 0.1347 & 0.3320 & 0.2027 & 0.0168 & 0.2027 & 0.2915 \\
Ultravox-v0.5 (LLaMA-3.1-8B) & 0.2216 & 0.1377 & 0.0973 & \textbf{0.4722} & \textbf{0.2813} & 0.0243 & \textbf{0.2813} & \textbf{0.3993} \\
Ultravox-v0.5 (LLaMA-3.2-1B) & 0.2209 & 0.1378 & 0.0936 & 0.4691 & 0.2751 & 0.0231 & 0.2751 & 0.3940 \\
\midrule

\rowcolor{OurMethodBg}
\multicolumn{9}{c}{\textit{Naturalistic Speech Dataset~\cite{openneurods004408}}} \\
\midrule
Audio-Flamingo-3 & \textbf{0.2446} & 0.1397 & 0.0923 & 0.3427 & 0.2234 & 0.0169 & 0.2234 & 0.3015 \\
Baichuan-Audio-Base & 0.1582 & 0.1193 & 0.0902 & 0.4610 & 0.2616 & 0.0193 & 0.2616 & 0.3717 \\
Baichuan-Audio-Instruct & 0.1669 & 0.1137 & 0.0774 & 0.5032 & 0.2915 & 0.0208 & 0.2915 & 0.4098 \\
GLM-4-Voice-9B & 0.2362 & \textbf{0.1773} & \textbf{0.1192} & 0.3314 & 0.2160 & 0.0151 & 0.2160 & 0.2881 \\
Granite-Speech-3.3-8B & 0.2185 & 0.1617 & 0.1155 & 0.4610 & 0.2637 & 0.0185 & 0.2637 & 0.3630 \\
Llama-3.1-8B-Omni & 0.1657 & 0.1217 & 0.0787 & 0.5562 & 0.3401 & 0.0261 & 0.3401 & 0.4595 \\
MiniCPM-o-2\_6 & 0.2045 & 0.1484 & 0.1053 & 0.3563 & 0.2145 & 0.0149 & 0.2145 & 0.2921 \\
Qwen2-Audio-7B & 0.2059 & 0.1615 & 0.1080 & 0.3561 & 0.2080 & 0.0143 & 0.2080 & 0.2863 \\
Qwen2-Audio-7B-Instruct & 0.2053 & 0.1467 & 0.1124 & 0.3649 & 0.2148 & 0.0147 & 0.2148 & 0.2982 \\
SpeechGPT-2.0-preview-7B & 0.2243 & 0.1564 & 0.1102 & 0.3924 & 0.2287 & 0.0171 & 0.2287 & 0.3239 \\
Ultravox-v0.5 (LLaMA-3.1-8B) & 0.1480 & 0.0815 & 0.0587 & \textbf{0.5696} & \textbf{0.3459} & \textbf{0.0266} & \textbf{0.3459} & \textbf{0.4664} \\
Ultravox-v0.5 (LLaMA-3.2-1B) & 0.1403 & 0.0857 & 0.0580 & 0.5682 & 0.3438 & 0.0260 & 0.3438 & 0.4633 \\
\bottomrule
\end{tabular}
}

\caption{Best-layer audio LLM--EEG similarity results on 2 datasets across 8 metrics. All corresponding permutation-test $p$-values are below 0.05. Best-layer is selected on the sentence-averaged curve.}
\label{tab:best_layer_d1_d2_stacked}
\end{table*}

\subsection{Experimental Setup}
\paragraph{Datasets.} 

We evaluate neural--model alignment using two naturalistic speech-listening electroencephalography (EEG) datasets. The first dataset is the sentence-segmented \emph{Alice in Wonderland} corpus \citep{brennan2019hierarchical_predictions}, which contains recordings from 49 subjects listening to Chapter~1 of \emph{Alice's Adventures in Wonderland} (84 sentences; 12.4\,min). Using the provided word- and sentence-level alignment tables, we segment the continuous EEG recordings into 84 sentence-locked epochs per subject and retain 60 scalp channels sampled at 500\,Hz after excluding auxiliary sensors. The second dataset is the Lalor Lab \emph{Naturalistic Speech} EEG dataset~\cite{openneurods004408}, comprising 736 sentence recordings from 19 subjects with 128 channels sampled at 512\,Hz, collected under a similar naturalistic speech-listening paradigm \citep{diliberto2015phoneme,broderick2018semantic}. For both datasets, continuous EEG signals are segmented into sentence-level epochs aligned with the corresponding audio stimuli.

\paragraph{Audio LLMs.}
We compare 12 Audio LLMs by extracting layer-wise hidden representations and aligning them to EEG.
Our model suite includes \textsc{Audio-Flamingo-3} \citep{goel2025audioflamingo3},
\textsc{Baichuan-Audio-Base} and \textsc{Baichuan-Audio-Instruct} \citep{li2025baichuanaudio},
\textsc{GLM-4-Voice-9B} \citep{zeng2024glm4voice},
\textsc{Granite-Speech-3.3-8B} \citep{saon2025granitespeech},
\textsc{Llama-3.1-8B-Omni} (via the \textsc{LLaMA-Omni} speech-interaction architecture) \citep{fang2024llamaomni},
\textsc{MiniCPM-o-2.6} \citep{openbmb2025minicpmo2626},
\textsc{Qwen2-Audio-7B} and \textsc{Qwen2-Audio-7B-Instruct} \citep{chu2024qwen2audio},
\textsc{SpeechGPT-2.0-preview-7B} \citep{openmoss2025speechgpt2previewpreviewpreview,zhang2023speechgpt},
and \textsc{Ultravox-v0.5} checkpoints built on \textsc{Llama-3.1-8B} and \textsc{Llama-3.2-1B} \citep{fixie2025ultravox}.

\paragraph{Implementation Details.}
We apply PCA with $k=20$ and assess significance using $n=500$ time-shuffle permutations \citep{jolliffe2002pca, nichols2002nonparametric}. For electrode-wise enriched features, we compute time-domain statistics using sliding windows of lengths $w\in\{3,5,9\}$ (in token-synchronous samples) and FFT-based band energies using window lengths $w\in\{8,16,32\}$. For prosody clustering, we set $K=4$ in $K$-means \citep{macqueen1967kmeans,lloyd1982kmeans}. For affect proxies, we set $(a_1,a_2,a_3)=(0.55,-0.25,-0.20)$. 
Full implementation and compute details are provided in Appendix~\ref{app:impl_details}.

\subsection{Audio LLM-Brain Similarity}
We evaluate 12 models on two datasets and report mean $\pm$ standard error across subjects. Table~\ref{tab:best_layer_d1_d2_stacked} summarizes best-layer results for eight similarity metrics (layer-wise trajectories in Appendix~\ref{sec:appendix_layerwise}). Across both datasets, model rankings are strongly metric-dependent: rank/geometry-based measures favor different models than dependence-based measures. On \textit{Alice in Wonderland}, \textsc{audio-flamingo-3} leads Pearson RSA, \textsc{glm-4-voice-9b} leads Spearman RSA, \textsc{granite-speech-3.3-8b} leads Kendall, and \textsc{Ultravox-v0.5} (\textsc{Llama-3.1-8B}) achieves the strongest dependence-based scores despite weaker rank-based alignment. The same pattern holds on \textit{Naturalistic Speech}: \textsc{audio-flamingo-3} again leads Pearson RSA, \textsc{glm-4-voice-9b} ranks highest under Spearman RSA and Kendall, and Ultravox-v0.5 attains the strongest dependence-based performance. Overall, these results show a consistent rank–dependence split, motivating the use of multiple complementary metrics for audio LLM–brain alignment.

\subsection{Electrode-wise Enriched Similarity}
The analyses above quantify EEG--model alignment at the sentence and layer levels, but they collapse spatial information across electrodes. To preserve sensor-level structure and examine spatially distributed alignment patterns, we perform an electrode-wise similarity analysis.

For each electrode, we construct an enriched EEG representation by concatenating complementary temporal and spectral features, including instantaneous amplitude, local dynamics, windowed statistics, signal power, and band-limited energy \citep{cohen2014analyzing,oppenheim2009dtsp,fulcher2013hctsa}. We then compute similarity independently for each electrode by correlating its enriched, token-aligned EEG sequence with the corresponding Audio LLM representations along the same temporal axis, rather than flattening across electrodes. This yields spatially resolved EEG--model similarity profiles that support subsequent topographic and time-resolved analyses; detailed feature definitions are provided in Appendix~\ref{app:enriched_eeg_features}.
\begin{figure}[t]
    \centering
    \includegraphics[width=1\linewidth]{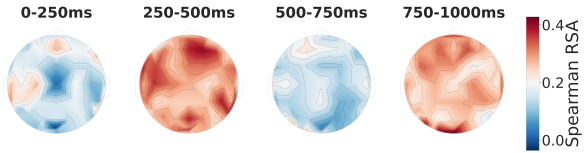}
\caption{Spatiotemporal dynamics of topographic similarity between Audio LLM and EEG signals.}

\label{fig:topo_windows}
\end{figure}

\begin{figure}[t]
    \centering
    \includegraphics[width=0.85 \linewidth]{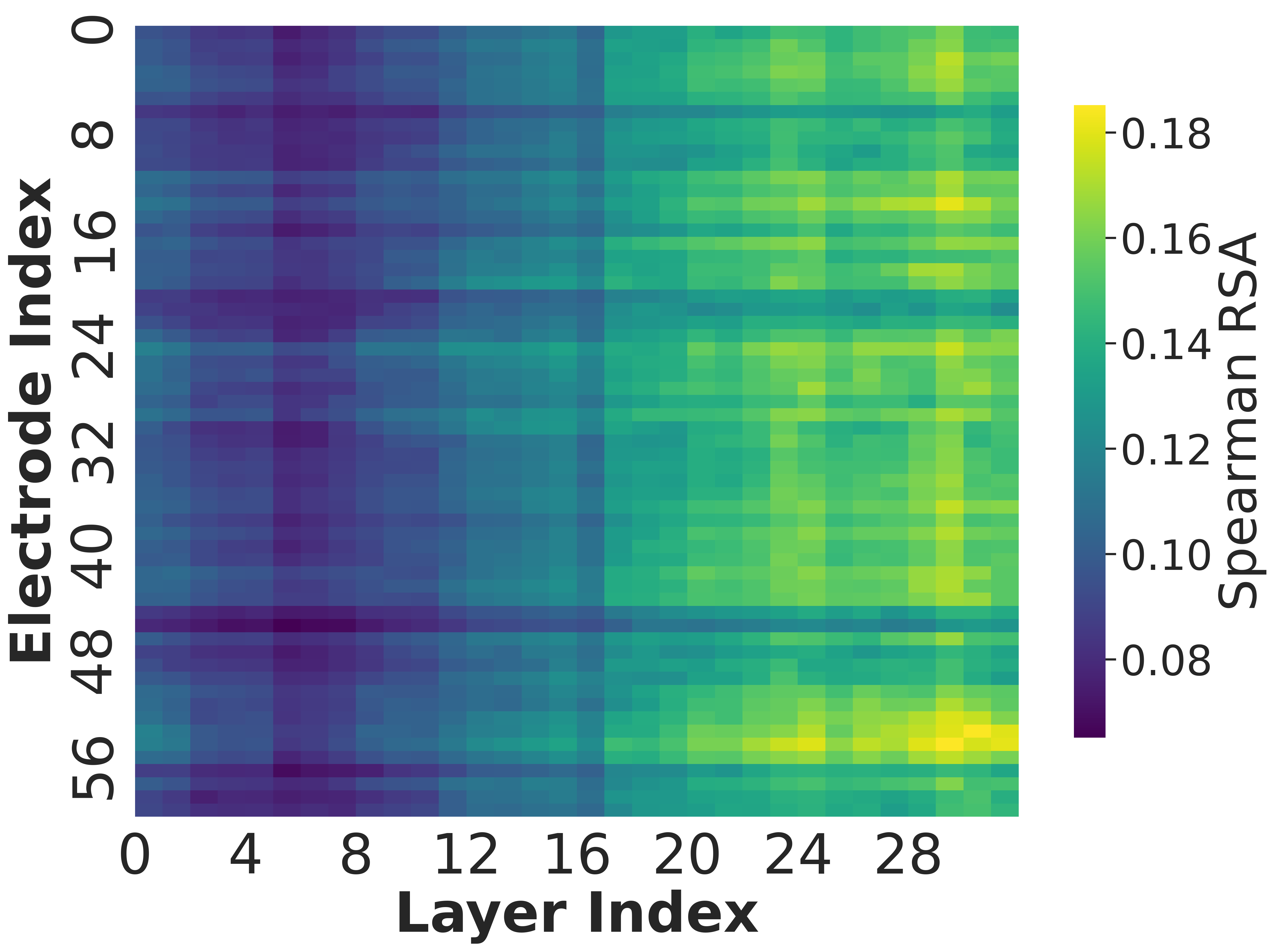}

\caption{The heatmap results between electrodes and audio LLM layers.}
\label{fig:heatmap}
\end{figure}

\subsubsection{Time-resolved Topographies}
To pinpoint when EEG–model geometric alignment emerges during sentence processing, we compute time-resolved RSA in four non-overlapping 250\,ms windows spanning 0–1000\,ms, following prior continuous-speech EEG studies \citep{lalor2010uninterrupted_speech,diliberto2015phoneme}; details of the time-window partitioning procedure are provided in Appendix~\ref{app:impl_details}. In each window, EEG signals aligned and resampled to the model time axis are compared with the corresponding model hidden states using electrode-wise Spearman RSA, and the resulting patterns are shown as scalp topographies in Figure~\ref{fig:topo_windows} and interpreted relative to the canonical N400 latency range \citep{kutas2011n400review}. Alignment is weakest at 0–250\,ms, peaks at 250–500\,ms, attenuates at 500–750\,ms, and partially re-emerges at 750–1000\,ms. The 250–500\,ms peak falls within the N400 range linked to higher-level language processing \citep{kutas1980senseless_sentences,kutas2011n400review} and exhibits a spatially distributed scalp pattern consistent with distributed networks supporting naturalistic speech comprehension \citep{hamilton2021parallel}.

\begin{figure}[t]
    \centering
    \includegraphics[width=1\linewidth]{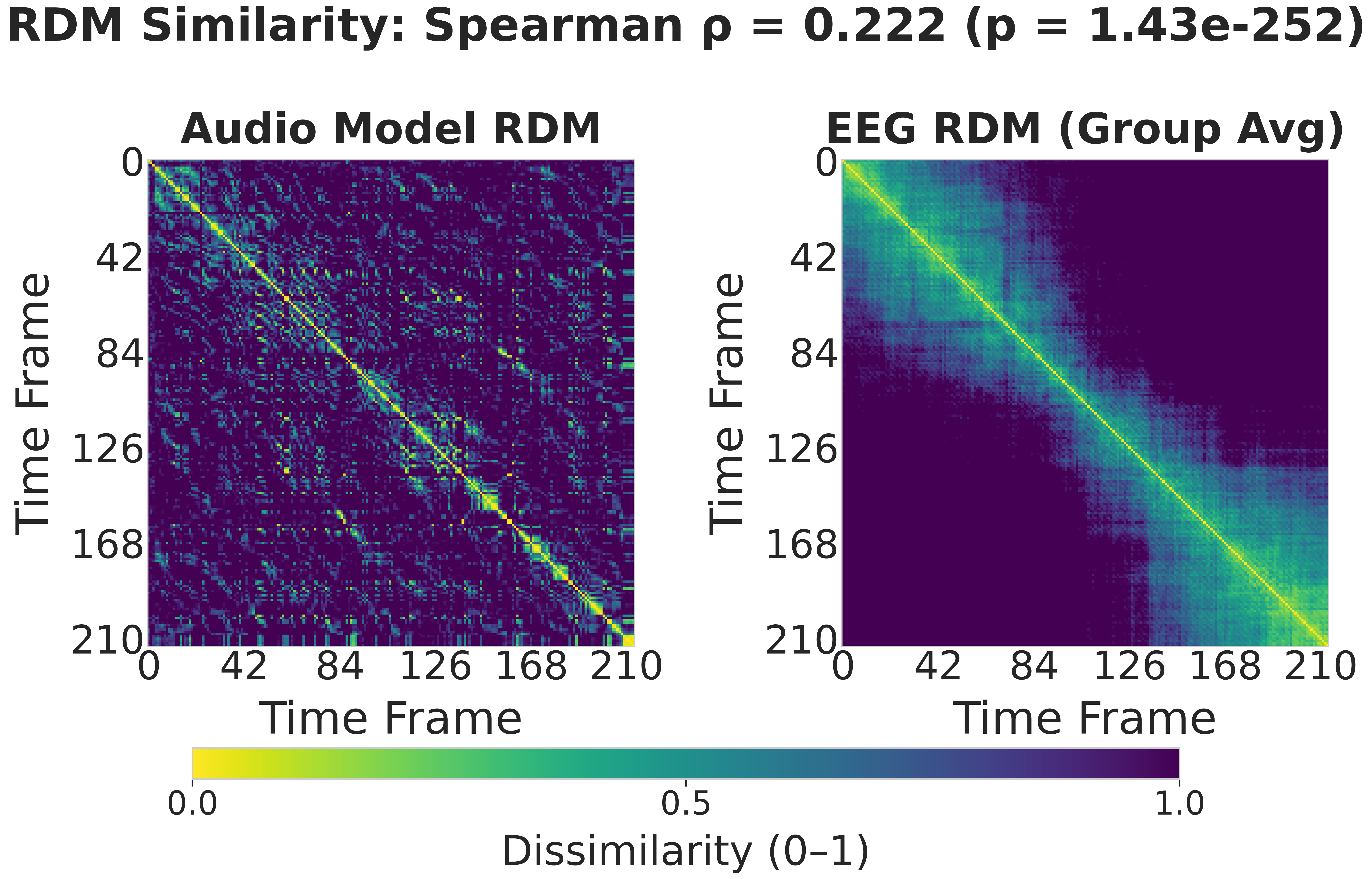}

\caption{Paired representational dissimilarity matrices for audio model and EEG.}
\label{fig:rdm_pair}
\end{figure}

\subsubsection{Heatmaps and RDM Visualizations}
To characterize the spatial and hierarchical structure of EEG–model alignment, Figure~\ref{fig:heatmap} shows weak similarity in early layers that increases toward mid-to-late layers, indicating depth-dependent geometric correspondence and non-uniformity across electrodes. Figure~\ref{fig:rdm_pair} further reveals modality-specific RDM structure: the audio--model RDM shows finer local variation, whereas the EEG RDM is smoother and more banded, consistent with temporally extended neural dynamics. Despite these differences, the strong, highly significant RDM similarity indicates a shared relative ordering of time-step dissimilarities, highlighting that EEG–model alignment reflects representational-geometry agreement rather than pointwise similarity. All sentence-level visualizations are provided in Appendix~\ref{sec:Supplementary Heatmaps and Paired RDM Visualizations}.

\subsection{Inter-subject Consistency}

\begin{figure}[t]
    \centering
    \includegraphics[width=0.8\linewidth]{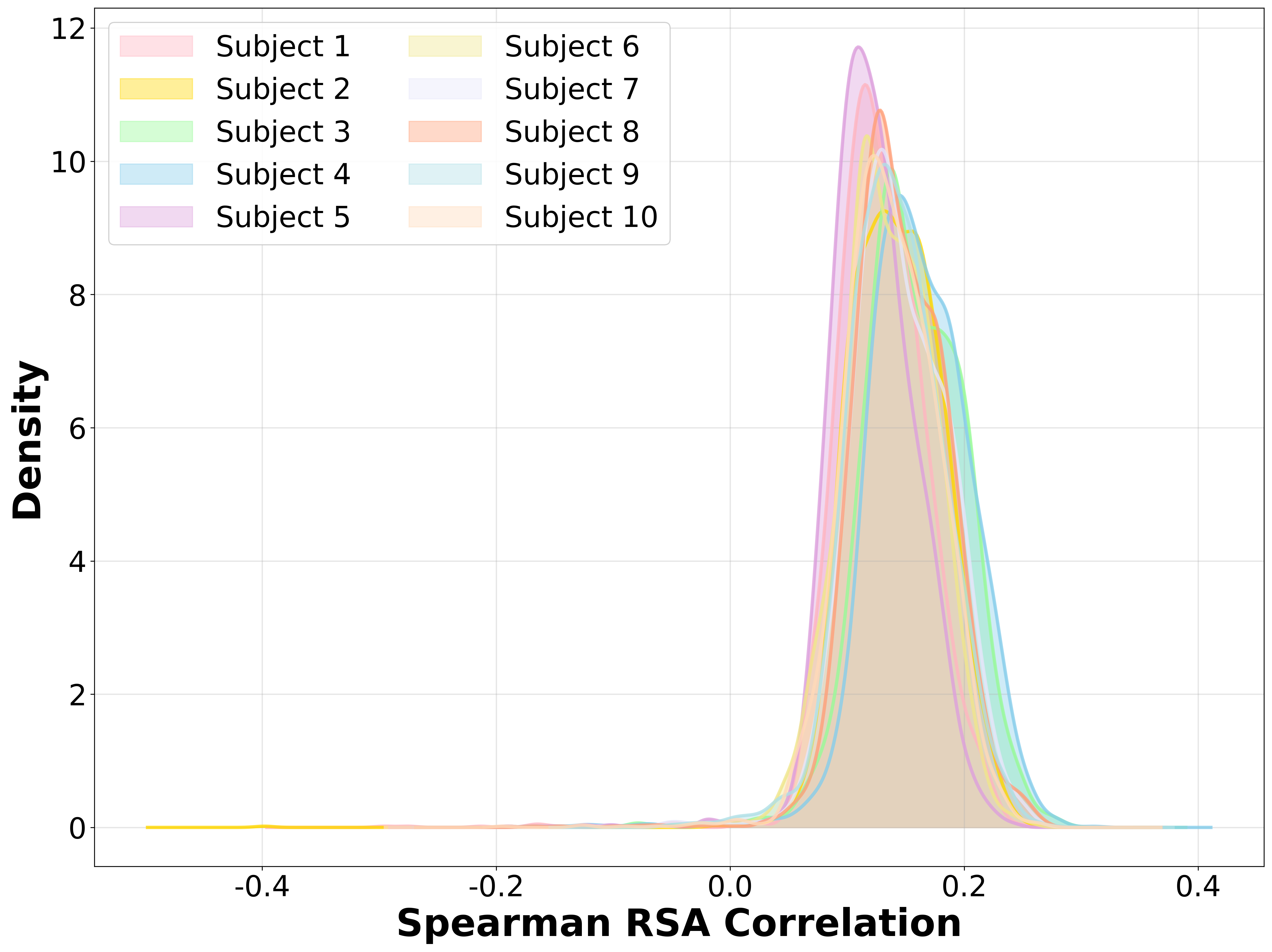}

\caption{Subject-wise distribution of best-layer Spearman RSA. }
\label{fig:subject_bar}
\end{figure}

To assess the stability of EEG--model correspondence across participants, we examine subject-wise distributions of best-layer Spearman RSA scores in Figure~\ref{fig:subject_bar}. Spearman RSA values are predominantly positive across subjects, with largely overlapping distributions and modest shifts in central tendency, indicating limited but non-negligible inter-subject variability.

\subsection{Emotion Partitioning}
\begin{figure}[t]
\centering
\includegraphics[width=1\columnwidth]{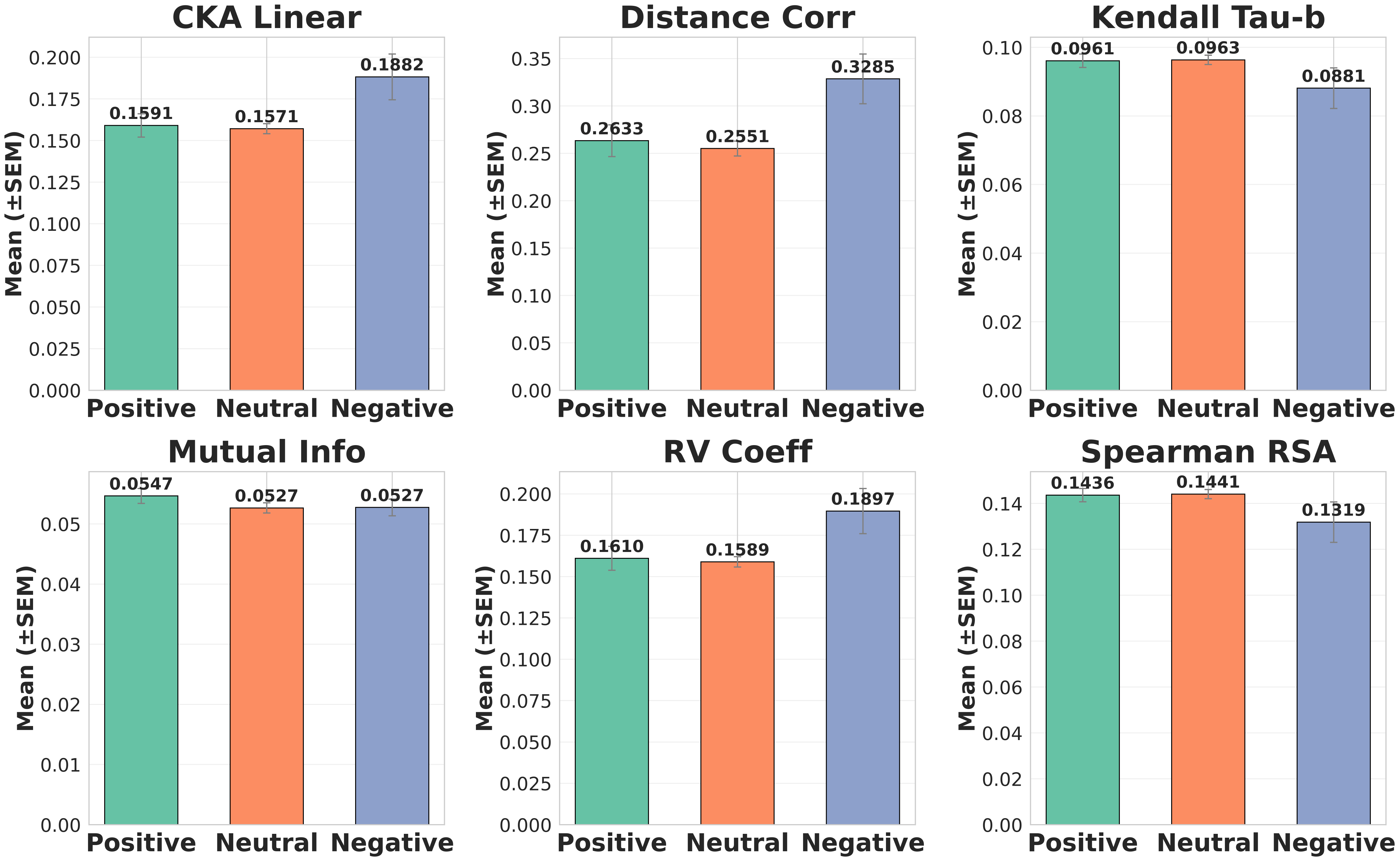}
\caption{EEG--Audio LLM similarity across different sentiment groups.}
\label{fig:sentiment_bar}
\end{figure}

Figure~\ref{fig:sentiment_bar} shows that affective prosody modulates EEG–model alignment in a metric-dependent manner. Geometry-sensitive rank measures, including Spearman RSA and Kendall’s $\tau_b$, are reduced for \textsc{Negative} stimuli, indicating weaker preservation of relative dissimilarity structure \citep{kriegeskorte2008rsa}. In contrast, dependence-based measures such as CKA, distance correlation, and RV are elevated for \textsc{Negative} prosody, reflecting stronger global covariance alignment \citep{kornblith2019cka,szekely2007dcor,escoufier1973traitement}. Mutual-information estimates remain largely stable across conditions, suggesting limited modulation at the local information level. Overall, negative prosody strengthens broad statistical dependence while disrupting fine-grained geometric alignment, consistent with enhanced processing of negative affective cues \citep{vuilleumier2005beware}.

\begin{figure}[t]
    \centering
    \includegraphics[width=1\linewidth]{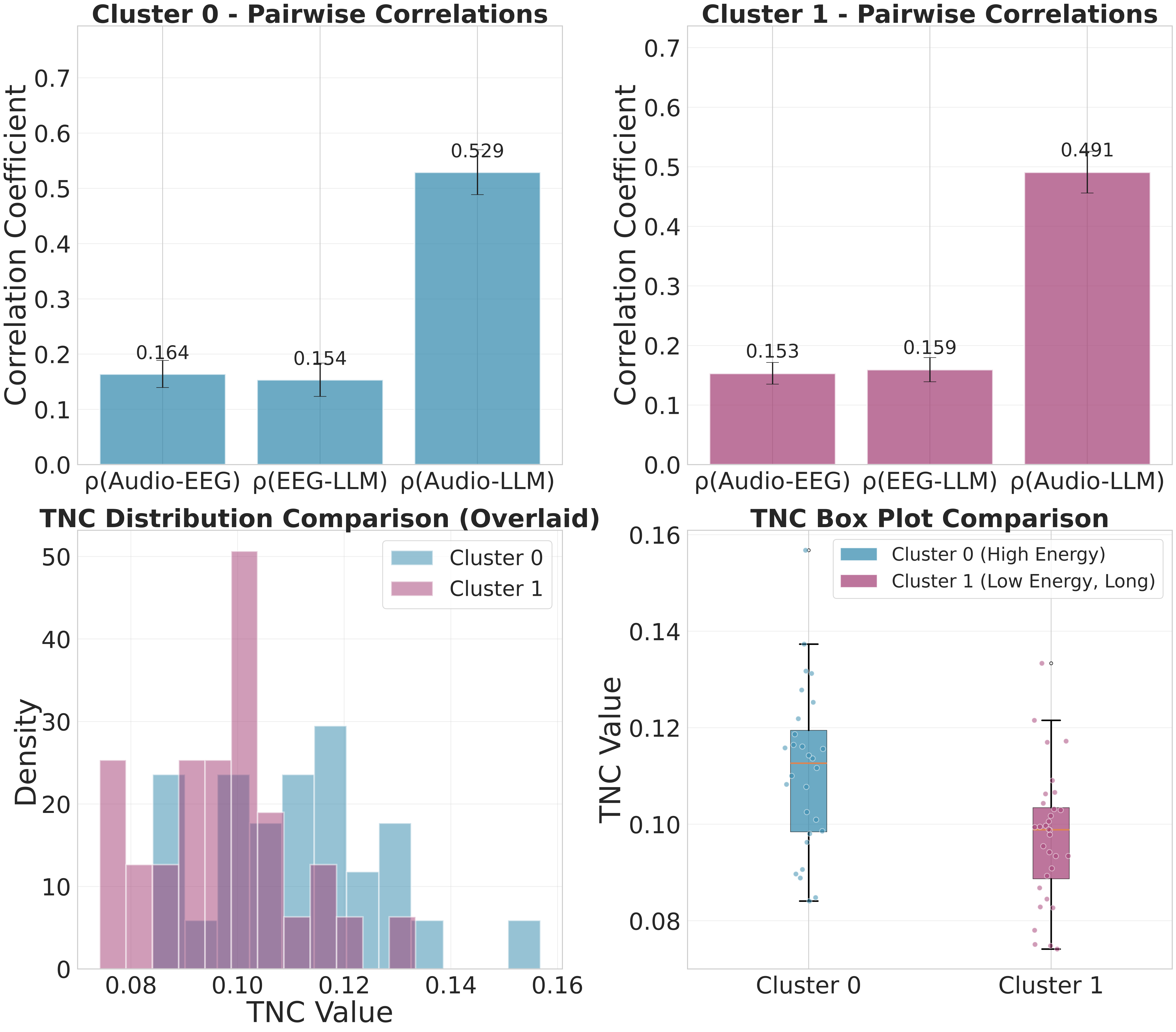}
\caption{TNC results grouped by prosody clusters. For the Audio LLM, we use representations from the final transformer layer.}
\label{fig:prosody_clusters}
\end{figure}

\subsection{Prosody Clustering and TNC Analysis}
TNC distributions within prosody clusters are shown in Figure~\ref{fig:prosody_clusters}. Across all clusters, Audio–Model correlations consistently exceed both Audio–EEG and EEG–Model, indicating that model representations more directly reflect acoustic neighborhood structure, while EEG exhibits weaker but relatively stable coupling to the stimulus. Prosody primarily modulates the \emph{dispersion} of TNC rather than its mean: high-energy clusters show greater variability, whereas low-energy or longer-duration clusters yield more concentrated distributions. These results suggest that prosodic conditions influence not only pairwise alignment strength but also how coherently EEG–model similarity is anchored to stimulus acoustics, consistent with evidence linking salient acoustic–rhythmic structure to distributed cortical dynamics \citep{poeppel2020rhythms,hamilton2021parallel}.

\section{Conclusion}

To investigate how Audio LLMs align with human neural processing during naturalistic speech listening, we conducted a neural study quantifying layer-wise correspondence with EEG across a comprehensive suite of 8 similarity metrics.
Our findings reveal pronounced depth-dependent alignment patterns and demonstrate that cross-model rankings depend strongly on the similarity metric, with rank- and geometry-based measures peaking at different layers than dependence-based measures. Prosody-aware analyses further uncover systematic condition effects, while time-resolved scalp topographies show stronger geometric alignment in the 250–500ms window, consistent with canonical N400-range dynamics. These results highlight the multifaceted nature of model–brain alignment and motivate future work on richer temporal modeling, larger and more diverse cohorts, and controlled manipulations to disentangle acoustic and linguistic contributions to neural correspondence.

\section*{Limitations}
While the observed EEG–Audio-LLM alignment is encouraging, several limitations warrant caution. Our benchmark is restricted to open-sourced Audio LLM checkpoints, which ensures reproducibility and layer-wise access but leaves open whether the observed rank–dependence dissociations and depth-wise trends generalize to large closed-source systems; extending evaluations to proprietary models, when intermediate representations are accessible, would help disentangle effects of scale versus architecture. In addition, sensor-space EEG is inherently limited by volume conduction and coarse anatomical specificity, suggesting that source-resolved analyses or complementary modalities such as MEG or fMRI could provide sharper localization of language-related activity. Nonetheless, EEG remains well suited to this setting due to its non-invasiveness, scalability, and millisecond temporal resolution, which enables precise characterization of when model–brain alignment emerges during naturalistic speech processing.

\section*{Ethics Statement}
This study analyzes previously collected, publicly available EEG datasets and open-source Audio LLM checkpoints. We do not collect new human-subject data, and we do not attempt to identify, re-contact, or infer personal attributes of any participant.

\paragraph{Human data and privacy.}
For the \textit{Alice in Wonderland} EEG dataset, participants in the original study provided written informed consent and the experiment received IRB approval (University of Michigan HSBS IRB \#HUM00081060).
For the naturalistic speech EEG dataset hosted on OpenNeuro (ds004408), data are shared for secondary analysis as de-identified human-subject recordings under OpenNeuro’s access terms; we follow these terms and use the data only for research purposes.

\paragraph{Data and code release.}
Our released artifacts contain analysis code, aggregate statistics, and derived figures, but do not include any personally identifying information. We follow the datasets’ respective licenses and attribution requirements when distributing derived results.

\paragraph{Potential downstream risks and mitigation.}
Model--brain alignment research can raise privacy concerns if misused for invasive inference. Our work focuses on representational similarity (RDM-based geometry) rather than reconstructing speech or decoding private content, and we report results primarily in aggregated form. We encourage future work to maintain strong governance (consent, de-identification, and access control) when extending alignment methods or applying them to more sensitive settings.

\bibliography{custom}

\appendix
\section*{Appendix}
\section{Implementation Details}
\label{app:impl_details}

All experiments were conducted on a single NVIDIA RTX A6000 GPU (48GB VRAM).

\subsection{Dimensionality reduction and RSA computation}
\paragraph{PCA dimensionality ($k=20$).}
To reduce computational cost and to match representational dimensionality across modalities, we apply PCA to both EEG-derived features and audio/model embeddings and retain $k=20$ principal components \citep{jolliffe2002pca}.
This choice is a compute--fidelity compromise: it (i) reduces the quadratic cost of RDM construction, (ii) mitigates noise and collinearity in high-dimensional embeddings, and (iii) enforces a comparable effective dimensionality across modalities for fair similarity scoring.

\paragraph{RDM construction (correlation distance) and RSA.}
We construct RDMs using correlation distance and compute RSA by correlating the upper-triangular vectorizations of the two RDMs \citep{kriegeskorte2008rsa}.

\subsection{Permutation-based significance testing}
For a given sentence $s$ (and a fixed layer/metric setting), let $r_{\mathrm{obs}}$ denote the observed similarity score.
We assess significance via time-shuffle permutation testing with $n=500$ permutations, which yields a practical balance between runtime and stability of the null estimate \citep{nichols2002nonparametric}.
Because permutation $p$-values are discrete, $n=500$ implies a minimum attainable resolution on the order of $1/(n+1)$, which is typically adequate for standard significance thresholds.

\paragraph{Null generation by breaking temporal correspondence.}
We generate the null distribution by randomly permuting the aligned time indices in one modality; for each permutation
$b\in\{1,\ldots,n\}$ we recompute the score, yielding $\{r_{\mathrm{perm}}^{(b)}\}_{b=1}^{n}$.
We use the standard one-sided permutation $p$-value estimator
\begin{equation}
p=\frac{\#\left\{r_{\mathrm{perm}}^{(b)}\ge r_{\mathrm{obs}}\right\}+1}{n+1}.
\end{equation}

\paragraph{GPU batching (16{,}384 per batch).}
Permutations are computed on GPU in FP32 and processed in batches of 16{,}384.
We choose this batch size because it is large enough to amortize kernel launch overhead and exploit GPU parallelism, while typically fitting within memory constraints; using a power-of-two batch size also tends to align well with GPU-friendly tensor shapes.

\subsection{Enriched EEG feature window lengths}
\paragraph{Time-domain sliding windows ($w\in\{3,5,9\}$).}
For local time-domain summary statistics, we use short windows $w\in\{3,5,9\}$ (in token-synchronous samples) to capture dynamics at multiple scales (very short / short / mid-range context) without oversmoothing transient changes.

\paragraph{FFT band-energy windows ($w\in\{8,16,32\}$).}
For FFT-based band energies, we use $w\in\{8,16,32\}$ to obtain more stable spectral estimates than ultra-short windows.
Power-of-two window sizes also enable efficient FFT computation \citep{cooley1965fft}, reducing runtime in large-scale sentence-by-sentence processing.

\subsection{Partition hyperparameters and feature sets}
\label{app:partition-details}

This appendix specifies the concrete values and feature sets used for sentence partitioning by affect and prosody.

\paragraph{Affect partition (valence threshold $\tau_v=0.45$ and proxy weights).}
We assign sentences to three affect groups using a symmetric threshold on the valence proxy:
positive if $\mathrm{valence}(s)>\tau_v$, negative if $\mathrm{valence}(s)<-\tau_v$, and neutral otherwise.
We set $\tau_v=0.45$ to separate mildly-to-strongly positive/negative sentences from a central neutral band without making the extreme groups too small for stable aggregation.
For the valence proxy, after $z$-scoring each contributing acoustic feature across sentences, we set
$(a_1,a_2,a_3)=(0.55,-0.25,-0.20)$ to encode a priori directional contributions:
positive weights emphasize features expected to increase the proxy score, while negative weights downweight features expected to contribute inversely.
Because all features are standardized, the relative magnitudes of these weights directly control the intended dominance among terms.

\paragraph{Prosody partition (13-D descriptor and $K$-means with $K=4$).}
For prosody clustering, we form a 13-dimensional descriptor vector for each sentence:
\[
\mathbf{p}_s \in \mathbb{R}^{13},
\]
where $\mathbf{p}_s$ concatenates 13 prosody descriptors: duration, energy\_mean, energy\_std, energy\_range, f0\_mean, f0\_std, f0\_range, f0\_slope, voiced\_ratio, energy\_diff\_mean, zcr\_mean, spectral\_centroid\_mean, spectral\_bandwidth\_mean.

We choose this compact set to cover complementary prosodic cues---timing/rhythm (duration), intensity dynamics (energy statistics and energy differences), intonation (F0 level/variability/contour), voicing/periodicity (voiced ratio and ZCR), and spectral-shape correlates of voice quality/articulation (spectral centroid/bandwidth)---using standard low-level descriptors implemented in openSMILE and related minimalistic voice feature sets \citep{eyben2010opensmile,eyben2016egemaps,rosenberg2009thesis}.

We $z$-score normalize each descriptor across sentences and run $K$-means with $K=4$ clusters \citep{macqueen1967kmeans,lloyd1982kmeans}.
To provide interpretable labels, we examine cluster centroids based on prosodic characteristics:
\begin{itemize}
\item \texttt{Cluster 0} (High Energy): 28 sentences with higher energy ($-28.6$ dB), medium duration (6.51s), and lower F0 (204.7 Hz);
\item \texttt{Cluster 1} (Low Energy, Long): 32 sentences with lower energy ($-33.9$ dB), longest duration (11.81s), and higher F0 (223.7 Hz);
\item \texttt{Cluster 2}: 1 sentence with extremely short duration (0.50s);
\item \texttt{Cluster 3}: 23 sentences with medium energy, shorter duration (4.70s), and higher F0.
\end{itemize}
For each cluster, we compute sentence-level TNC (Eq.~\ref{eq:tnc}) and report the cluster-wise mean and standard deviation.
In the main text, we focus our analysis on \texttt{Cluster 0} and \texttt{Cluster 1} for three reasons: (i) \texttt{Cluster 2} contains only a single sentence, which is insufficient for reliable statistical inference; (ii) \texttt{Cluster 0} and \texttt{Cluster 1} exhibit the most distinctive and interpretable prosodic contrast (high vs.\ low energy); and (iii) together they account for 60 out of 84 sentences (71.4\%), providing sufficient statistical power for downstream analysis.

\subsection{Defining millisecond windows on the token-synchronous axis}
Our RSA computations operate on a token-synchronous time grid of length $T_a$ (main text), so we require a principled way to express fixed real-time windows (in ms) on this discrete index axis.
For each sentence $s$ with total duration $D_s$ milliseconds, we treat the token-synchronous grid as uniformly spaced over $[0,D_s)$ and assign each step $t\in\{1,\ldots,T_a\}$ the timestamp
\begin{equation}
\tau_s(t) \;=\; (t-1)\cdot \frac{D_s}{T_a}\quad\text{(milliseconds)}.
\end{equation}
Given a real-time window $[a,b)$\,ms, we select the corresponding set of token indices as
\begin{equation}
\small
\mathcal{T}_{s}[a,b) \;=\;\{\,t\in\{1,\ldots,T_a\}\;:\; a \le \tau_s(t) < b\,\}.
\end{equation}
We compute window-specific RSA using only the token-aligned samples indexed by $\mathcal{T}_{s}[a,b)$ (for both EEG and model states).
In the main analysis, we use four non-overlapping windows: $[0,250)$, $[250,500)$, $[500,750)$, and $[750,1000)$\,ms, and truncate to the available range via $[a,\min(b,D_s))$ when $D_s<b$.
If a sentence yields fewer than two token-synchronous samples in a window (i.e., $|\mathcal{T}_{s}[a,b)|<2$), we omit that sentence-window instance from RSA for numerical stability.

\section{Metric Definitions}
\label{app:metric-definitions}

This appendix provides formal definitions of the complementary dependence measures reported in Sec.~\ref{sec:similarity}.
We follow the notation in Sec.~\ref{sec:similarity}: for each sentence $s$, EEG and layer-$l$ Audio LLM states are represented as
\(\mathbf{U}^{\mathrm{eeg}}_s=[\mathbf{u}^{\mathrm{eeg}}_{s,1},\ldots,\mathbf{u}^{\mathrm{eeg}}_{s,T_a}]^\top\in\mathbb{R}^{T_a\times d_{\mathrm{eeg}}}\) and
\(\mathbf{V}^{\mathrm{llm},l}_s=[\mathbf{v}^{\mathrm{llm},l}_{s,1},\ldots,\mathbf{v}^{\mathrm{llm},l}_{s,T_a}]^\top\in\mathbb{R}^{T_a\times d_{\mathrm{llm}}}\),
where \(T_a\) is the aligned sequence length and \(d_{\mathrm{eeg}},d_{\mathrm{llm}}\) are feature dimensions after any PCA.

\subsection{Kendall's $\tau_b$}
Kendall's $\tau_b$ measures tie-corrected rank association \citep{kendall1945tiesties}.
Given two real-valued vectors \(\mathbf{a},\mathbf{b}\in\mathbb{R}^{n}\) (in our setting,
\(\mathbf{a}=\mathbf{r}^{\mathrm{eeg}}_s\) and \(\mathbf{b}=\mathbf{r}^{\mathrm{llm},l}_s\), the vectorized upper-triangular RDMs),
\begin{equation}
\tau_b=\frac{C-D}{\sqrt{(C+D+T_{\mathbf{a}})(C+D+T_{\mathbf{b}})}},
\end{equation}
where \(C\) and \(D\) are the numbers of concordant and discordant pairs, and
\(T_{\mathbf{a}}\) and \(T_{\mathbf{b}}\) count ties in \(\mathbf{a}\) and \(\mathbf{b}\), respectively.

\subsection{Distance Correlation (dCor)}
Distance correlation is zero if and only if independence under mild conditions \citep{szekely2007dcor}.
Let \(a_{ij}=\|\mathbf{u}^{\mathrm{eeg}}_{s,i}-\mathbf{u}^{\mathrm{eeg}}_{s,j}\|_2\) and
\(b_{ij}=\|\mathbf{v}^{\mathrm{llm},l}_{s,i}-\mathbf{v}^{\mathrm{llm},l}_{s,j}\|_2\) be pairwise Euclidean
distances across aligned time indices \(1\le i,j\le T_a\).
Define the doubly-centered distance matrices:
\begin{equation}
\begin{aligned}
\tilde{a}_{ij}
&= a_{ij}-\bar{a}_{i\cdot}-\bar{a}_{\cdot j}+\bar{a}_{\cdot\cdot}, \\
\tilde{b}_{ij}
&= b_{ij}-\bar{b}_{i\cdot}-\bar{b}_{\cdot j}+\bar{b}_{\cdot\cdot}.
\end{aligned}
\end{equation}

where \(\bar{a}_{i\cdot}=\frac{1}{T_a}\sum_{j}a_{ij}\), \(\bar{a}_{\cdot j}=\frac{1}{T_a}\sum_{i}a_{ij}\), and
\(\bar{a}_{\cdot\cdot}=\frac{1}{T_a^2}\sum_{i,j}a_{ij}\) (similarly for \(b\)).
The squared distance covariance/variance are
\begin{equation}
\small
\begin{aligned}
\mathrm{dCov}^2(\mathbf{U}^{\mathrm{eeg}}_s,\mathbf{V}&^{\mathrm{llm},l}_s)
= \frac{1}{T_a^2}\sum_{i=1}^{T_a}\sum_{j=1}^{T_a}\tilde{a}_{ij}\tilde{b}_{ij},\\
\mathrm{dVar}^2(\mathbf{U}^{\mathrm{eeg}}_s) &= \mathrm{dCov}^2(\mathbf{U}^{\mathrm{eeg}}_s,\mathbf{U}^{\mathrm{eeg}}_s),\\
\mathrm{dVar}^2(\mathbf{V}^{\mathrm{llm},l}_s) &= \mathrm{dCov}^2(\mathbf{V}^{\mathrm{llm},l}_s,\mathbf{V}^{\mathrm{llm},l}_s).
\end{aligned}
\end{equation}
and distance correlation is
\begin{equation}
\begin{aligned}
\small
&\mathrm{dCor}(\mathbf{U}^{\mathrm{eeg}}_s,\mathbf{V}^{\mathrm{llm},l}_s)\\&=
\frac{\mathrm{dCov}(\mathbf{U}^{\mathrm{eeg}}_s,\mathbf{V}^{\mathrm{llm},l}_s)}
{\sqrt{\mathrm{dVar}(\mathbf{U}^{\mathrm{eeg}}_s)\,\mathrm{dVar}(\mathbf{V}^{\mathrm{llm},l}_s)}}.
\end{aligned}
\end{equation}

\subsection{RV Coefficient}
For column-centered matrices \(\mathbf{U}^{\mathrm{eeg}}_s\in\mathbb{R}^{T_a\times d_{\mathrm{eeg}}}\) and
\(\mathbf{V}^{\mathrm{llm},l}_s\in\mathbb{R}^{T_a\times d_{\mathrm{llm}}}\),
\begin{equation}
\small
\begin{aligned}
&\mathrm{RV}\!\left(\mathbf{U}^{\mathrm{eeg}}_s,\mathbf{V}^{\mathrm{llm},l}_s\right)
\\&=
\frac{
\mathrm{tr}\!\left(
(\mathbf{U}^{\mathrm{eeg}}_s)^\top
\mathbf{V}^{\mathrm{llm},l}_s
(\mathbf{V}^{\mathrm{llm},l}_s)^\top
\mathbf{U}^{\mathrm{eeg}}_s
\right)
}{
\sqrt{
\mathrm{tr}\!\left(\big((\mathbf{U}^{\mathrm{eeg}}_s)^\top\mathbf{U}^{\mathrm{eeg}}_s\big)^2\right)
\,
\mathrm{tr}\!\left(\big((\mathbf{V}^{\mathrm{llm},l}_s)^\top\mathbf{V}^{\mathrm{llm},l}_s\big)^2\right)
}
}.
\end{aligned}
\end{equation}

\subsection{Gaussian Mutual-Information Proxy}
Under a bivariate Gaussian approximation with Pearson correlation \(r_{\mathrm{flat}}\), mutual information can be written as \citep{cover2006it}:
\begin{equation}
I=-\frac{1}{2}\log\!\left(1-r_{\mathrm{flat}}^2\right).
\end{equation}
In our implementation,
\(r_{\mathrm{flat}}=\mathrm{Pearson}\!\left(\operatorname{vec}(\mathbf{U}^{\mathrm{eeg}}_s),\,\operatorname{vec}(\mathbf{V}^{\mathrm{llm},l}_s)\right)\),
so this proxy increases monotonically with shared variance.

\subsection{Centered Kernel Alignment (CKA)}
CKA measures similarity between representations using centered Gram matrices \citep{kornblith2019cka}.
Let \(K=\kappa(\mathbf{U}^{\mathrm{eeg}}_s,\mathbf{U}^{\mathrm{eeg}}_s)\in\mathbb{R}^{T_a\times T_a}\) and
\(L=\kappa(\mathbf{V}^{\mathrm{llm},l}_s,\mathbf{V}^{\mathrm{llm},l}_s)\in\mathbb{R}^{T_a\times T_a}\)
denote Gram matrices (e.g., linear kernel \(K=\mathbf{U}\mathbf{U}^\top\), \(L=\mathbf{V}\mathbf{V}^\top\), or RBF kernels),
and let \(H=I-\frac{1}{T_a}\mathbf{1}\mathbf{1}^\top\) be the centering matrix. Define \(K_c=HKH\) and \(L_c=HLH\). Then
\begin{equation}
\small
\mathrm{CKA}(\mathbf{U}^{\mathrm{eeg}}_s,\mathbf{V}^{\mathrm{llm},l}_s)
=\frac{\mathrm{tr}(K_c L_c)}{\sqrt{\mathrm{tr}(K_c K_c)\,\mathrm{tr}(L_c L_c)}}.
\end{equation}

\section{Theoretical Properties of Tri-modal Neighborhood Consistency (TNC)}
\label{app:tnc-theory}

This appendix provides formal properties of the Tri-modal Neighborhood Consistency (TNC) defined in Eq.~\ref{eq:tnc}.
Throughout, we follow the main-text notation: for sentence $s$ and Audio LLM layer $l$ we compute Spearman RSA correlations
$\rho^{\mathrm{ac,eeg}}_s$, $\rho^{\mathrm{eeg,llm},l}_s$, and $\rho^{\mathrm{ac,llm},l}_s$ from vectorized RDMs.

\subsection{Tri-modal RSA objects and a compact matrix view}

Let the token-synchronous acoustic feature sequence for sentence $s$ be
$\mathbf{U}^{\mathrm{ac}}_s=[\mathbf{u}^{\mathrm{ac}}_{s,1},\ldots,\mathbf{u}^{\mathrm{ac}}_{s,T_a}]^\top\in\mathbb{R}^{T_a\times d_{\mathrm{ac}}}$,
the aligned EEG be $\mathbf{U}^{\mathrm{eeg}}_s\in\mathbb{R}^{T_a\times C}$ (main text),
and the layer-$l$ Audio LLM representation be $\mathbf{V}^{\mathrm{llm},l}_s\in\mathbb{R}^{T_a\times D}$ (main text).
Define correlation-distance RDMs as in Eq.~\ref{eq:rdm}:
\begin{equation}
\begin{aligned}
&\mathrm{RDM}^{m}_s(i,j)=1-\mathrm{corr}\!\left(\mathbf{z}^{m}_{s,i},\mathbf{z}^{m}_{s,j}\right),\\
&\qquad m\in\{\mathrm{ac,eeg,llm},l\},
\end{aligned}
\end{equation}
where $\mathbf{z}^{\mathrm{ac}}_{s,i}:=\mathbf{u}^{\mathrm{ac}}_{s,i}$, $\mathbf{z}^{\mathrm{eeg}}_{s,i}:=\mathbf{u}^{\mathrm{eeg}}_{s,i}$,
and $\mathbf{z}^{\mathrm{llm},l}_{s,i}:=\mathbf{v}^{\mathrm{llm},l}_{s,i}$.
Vectorize the strictly upper-triangular entries:
\begin{equation}
\begin{aligned}
&\mathbf{r}^{m}_s := \operatorname{vec}_{\triangle}\!\left(\mathrm{RDM}^{m}_s\right)\in\mathbb{R}^{M_s},\\
&\qquad M_s=\frac{T_a(T_a-1)}{2}.
\end{aligned}
\end{equation}
Spearman RSA correlations in the main text are
\begin{equation}
\begin{aligned}
&\rho^{m,n}_s \;=\; \mathrm{Spearman}\!\left(\mathbf{r}^{m}_s,\mathbf{r}^{n}_s\right)
\;\\&=\; \mathrm{Pearson}\!\left(\operatorname{rank}(\mathbf{r}^{m}_s),\,\operatorname{rank}(\mathbf{r}^{n}_s)\right),
\label{eq:spearman_as_pearson_on_ranks}
\end{aligned}
\end{equation}
where $\operatorname{rank}(\cdot)$ is the componentwise rank transform.

Define the (Spearman) tri-modal correlation matrix for sentence $s$ and layer $l$:
\begin{equation}
\mathbf{R}^{\,l}_s \;=\;
\begin{bmatrix}
1 & \rho^{\mathrm{ac,eeg}}_s & \rho^{\mathrm{ac,llm},l}_s\\
\rho^{\mathrm{ac,eeg}}_s & 1 & \rho^{\mathrm{eeg,llm},l}_s\\
\rho^{\mathrm{ac,llm},l}_s & \rho^{\mathrm{eeg,llm},l}_s & 1
\end{bmatrix}.
\label{eq:tri_corr_matrix}
\end{equation}
Then TNC in Eq.~\ref{eq:tnc} can be rewritten as a normalized Frobenius-energy of off-diagonal coherences:
\begin{equation}
\small
\begin{aligned}
&\mathrm{TNC}^{\,l}_s
\\&=\frac{1}{3}\Big(\big[\rho^{\mathrm{ac,eeg}}_s\big]^2+\big[\rho^{\mathrm{eeg,llm},l}_s\big]^2+\big[\rho^{\mathrm{ac,llm},l}_s\big]^2\Big)
\\&=\frac{\|\mathbf{R}^{\,l}_s\|_F^2-3}{6}.
\label{eq:tnc_frobenius}
\end{aligned}
\end{equation}
\paragraph{Proof.}
$\|\mathbf{R}^{\,l}_s\|_F^2=\sum_{i,j}(\mathbf{R}^{\,l}_s)_{ij}^2
=3+2\!\left(\big[\rho^{\mathrm{ac,eeg}}_s\big]^2+\big[\rho^{\mathrm{eeg,llm},l}_s\big]^2+\big[\rho^{\mathrm{ac,llm},l}_s\big]^2\right)$,
which yields Eq.~\ref{eq:tnc_frobenius}. 

Eq.~\ref{eq:tnc_frobenius} makes explicit that $\mathrm{TNC}^{\,l}_s$ measures \emph{overall tri-modal coherence} (the total squared off-diagonal agreement) rather than any single pairwise alignment.

\subsection{Basic guarantees: boundedness and ``high TNC $\Rightarrow$ all pairs high''}

\begin{proposition}[Range]
\label{prop:range}
For all $s,l$, $\mathrm{TNC}^{\,l}_s\in[0,1]$.
\end{proposition}
\paragraph{Proof.}
Each Spearman correlation satisfies $\rho^{m,n}_s\in[-1,1]$, hence $\big[\rho^{m,n}_s\big]^2\in[0,1]$.
TNC is the average of three numbers in $[0,1]$, therefore lies in $[0,1]$. 

\begin{proposition}[Near-1 TNC forces all three pairwise RSA scores to be near-1]
\label{prop:near1}
Fix $\delta\in[0,1]$. If $\mathrm{TNC}^{\,l}_s \ge 1-\delta$, then
\begin{equation}
\begin{aligned}
\big[\rho^{\mathrm{ac,eeg}}_s\big]^2 \ge 1-3\delta,\\\qquad
\big[\rho^{\mathrm{eeg,llm},l}_s\big]^2 \ge 1-3\delta,\\\qquad
\big[\rho^{\mathrm{ac,llm},l}_s\big]^2 \ge 1-3\delta,
\end{aligned}
\end{equation}
equivalently, $|\rho^{\cdot,\cdot}_s|\ge \sqrt{1-3\delta}$ for all three modality pairs.
\end{proposition}
\paragraph{Proof.}
Let $a_1,a_2,a_3\in[0,1]$ be the three squared correlations.
If $\frac{1}{3}(a_1+a_2+a_3)\ge 1-\delta$, then for any $i$,
$a_i \ge 3(1-\delta)-(a_j+a_k)\ge 3(1-\delta)-2 = 1-3\delta$ because $a_j,a_k\le 1$. 

Proposition~\ref{prop:near1} formalizes the statement that \emph{high} TNC (close to 1) is attainable only when \emph{all three}
pairwise geometries are simultaneously consistent.

\subsection{Conservativeness under modality-specific noise (attenuation)}

We now formalize why TNC is conservative when one modality is noisy.
Because Spearman is Pearson on ranks (Eq.~\ref{eq:spearman_as_pearson_on_ranks}), we analyze correlations after the rank transform.

Let
\begin{equation}
\begin{aligned}
&\tilde{\mathbf{r}}^{m}_s := \operatorname{rank}(\mathbf{r}^{m}_s)\in\mathbb{R}^{M_s},\\&
m\in\{\mathrm{ac,eeg,llm},l\}.
\end{aligned}
\end{equation}
Assume a classical measurement-error (additive noise) model on these rank vectors:
\begin{equation}
\tilde{\mathbf{r}}^{m}_s = \mathbf{z}_s + \boldsymbol{\epsilon}^{m}_s,
\label{eq:additive_noise_model}
\end{equation}
where $\mathbf{z}_s\in\mathbb{R}^{M_s}$ is the latent (shared) within-sentence geometry, and
$\boldsymbol{\epsilon}^{m}_s$ is modality-specific noise with
$\mathbb{E}[\boldsymbol{\epsilon}^{m}_s]=\mathbf{0}$,
$\mathrm{Cov}(\boldsymbol{\epsilon}^{m}_s,\boldsymbol{\epsilon}^{n}_s)=\mathbf{0}$ for $m\neq n$,
and $\mathrm{Cov}(\mathbf{z}_s,\boldsymbol{\epsilon}^{m}_s)=\mathbf{0}$.

Let $\sigma_z^2:=\mathrm{Var}(z_{s,q})$ and $\sigma_m^2:=\mathrm{Var}(\epsilon^m_{s,q})$ denote per-component variances
(assuming stationarity across components $q\in\{1,\ldots,M_s\}$ for notational simplicity).
Then for any two modalities $m\neq n$,
\begin{equation}
\begin{aligned}
\mathbb{E}\!\left[\rho^{m,n}_s\right]
=\mathbb{E}\!\left[\mathrm{Pearson}(\tilde{\mathbf{r}}^{m}_s,\tilde{\mathbf{r}}^{n}_s)\right]
\\\approx
\frac{\sigma_z^2}{\sqrt{(\sigma_z^2+\sigma_m^2)(\sigma_z^2+\sigma_n^2)}},
\label{eq:attenuation}
\end{aligned}
\end{equation}
i.e., the expected correlation is \emph{attenuated} by modality-specific noise.
Consequently,
\begin{equation}
\small
\begin{aligned}
&\mathbb{E}\!\left[\mathrm{TNC}^{\,l}_s\right]
\\&\approx \frac{1}{3}\sum_{(m,n)\in\mathcal{P}}
\left(
\frac{\sigma_z^2}{\sqrt{(\sigma_z^2+\sigma_m^2)(\sigma_z^2+\sigma_n^2)}}
\right)^{\!2},
\label{eq:tnc_expected_noise}
\end{aligned}
\end{equation}
where $\mathcal{P}=\{(\mathrm{ac,eeg}),(\mathrm{eeg,llm},l),(\mathrm{ac,llm},l)\}$.

\paragraph{Implication (conservative behavior).}
If \emph{any} modality has large noise variance (e.g., $\sigma_{\mathrm{eeg}}^2\gg\sigma_z^2$),
then both $\rho^{\mathrm{ac,eeg}}_s$ and $\rho^{\mathrm{eeg,llm},l}_s$ are strongly attenuated by Eq.~\ref{eq:attenuation},
and thus $\mathrm{TNC}^{\,l}_s$ is suppressed through two of its three terms.
In contrast, a single pairwise score such as $\rho^{\mathrm{ac,llm},l}_s$ could remain high even when EEG is noisy.
Therefore, averaging \emph{three} squared correlations makes TNC a conservative estimate of genuine tri-modal correspondence.

\subsection{Mitigating indirect/shared confounds: dilution of pair-specific effects}

We next formalize why TNC reduces inflation from \emph{pair-specific} confounds.

Consider a decomposition on rank-RDM vectors:
\begin{equation}
\begin{aligned}
&\tilde{\mathbf{r}}^{\mathrm{ac}}_s = \mathbf{z}_s + \mathbf{q}_s + \boldsymbol{\epsilon}^{\mathrm{ac}}_s,\\
&\tilde{\mathbf{r}}^{\mathrm{llm},l}_s = \mathbf{z}_s + \mathbf{q}_s + \boldsymbol{\epsilon}^{\mathrm{llm},l}_s,\\
&\tilde{\mathbf{r}}^{\mathrm{eeg}}_s = \mathbf{z}_s + \boldsymbol{\epsilon}^{\mathrm{eeg}}_s,
\label{eq:pair_confound_model}
\end{aligned}
\end{equation}
where $\mathbf{q}_s$ is a \emph{shared confound} that influences acoustics and the model, but is not present in EEG.
Assume $\mathbf{z}_s$, $\mathbf{q}_s$, and all noises are mutually uncorrelated, zero-mean, with variances
$\sigma_z^2$, $\sigma_q^2$, $\sigma_{\mathrm{ac}}^2$, $\sigma_{\mathrm{llm}}^2$, $\sigma_{\mathrm{eeg}}^2$.

Then the \emph{confounded} pairwise correlation is
\begin{equation}
\small
\rho^{\mathrm{ac,llm},l}_s
\approx
\frac{\sigma_z^2+\sigma_q^2}{\sqrt{(\sigma_z^2+\sigma_q^2+\sigma_{\mathrm{ac}}^2)(\sigma_z^2+\sigma_q^2+\sigma_{\mathrm{llm}}^2)}},
\label{eq:confounded_pair}
\end{equation}
which can be large even if $\sigma_z^2$ is small (inflation driven by $\sigma_q^2$).
However, correlations involving EEG become
\begin{equation}
\small
\begin{aligned}
&\rho^{\mathrm{ac,eeg}}_s
\approx
\frac{\sigma_z^2}{\sqrt{(\sigma_z^2+\sigma_q^2+\sigma_{\mathrm{ac}}^2)(\sigma_z^2+\sigma_{\mathrm{eeg}}^2)}},\\
&\rho^{\mathrm{eeg,llm},l}_s
\approx
\frac{\sigma_z^2}{\sqrt{(\sigma_z^2+\sigma_{\mathrm{eeg}}^2)(\sigma_z^2+\sigma_q^2+\sigma_{\mathrm{llm}}^2)}}.
\label{eq:unconfounded_pairs}
\end{aligned}
\end{equation}
Notably, $\sigma_q^2$ appears only in the \emph{denominators} of Eq.~\ref{eq:unconfounded_pairs}, which further \emph{reduces}
$\rho^{\mathrm{ac,eeg}}_s$ and $\rho^{\mathrm{eeg,llm},l}_s$ as $\sigma_q^2$ grows.

\begin{proposition}[Pair-specific confounds inflate at most one term of TNC]
\label{prop:pair_confound_dilution}
Under the model in Eq.~\ref{eq:pair_confound_model}, increasing $\sigma_q^2$ can increase $\rho^{\mathrm{ac,llm},l}_s$ (Eq.~\ref{eq:confounded_pair})
but cannot increase $\rho^{\mathrm{ac,eeg}}_s$ or $\rho^{\mathrm{eeg,llm},l}_s$ (Eq.~\ref{eq:unconfounded_pairs});
indeed it typically decreases them. Consequently, any inflation driven by $\mathbf{q}_s$ contributes to at most one of the three squared
terms in $\mathrm{TNC}^{\,l}_s$.
\end{proposition}
\paragraph{Proof.}
Eq.~\ref{eq:unconfounded_pairs} shows $\sigma_q^2$ enters only inside a denominator term of the form
$\sqrt{\sigma_z^2+\sigma_q^2+\cdot}$, hence increasing $\sigma_q^2$ weakly decreases the corresponding fraction.
Meanwhile Eq.~\ref{eq:confounded_pair} contains $\sigma_q^2$ in both numerator and denominator, allowing potential increases.
TNC is the average of the three squared correlations (Eq.~\ref{eq:tnc}); thus $\mathbf{q}_s$ can inflate at most one addend. 

\paragraph{A sharp ``dilution'' special case.}
If $\sigma_z^2\approx 0$ (no true tri-modal shared geometry) and $\sigma_q^2$ dominates Eq.~\ref{eq:confounded_pair} while EEG is independent,
then $\rho^{\mathrm{ac,eeg}}_s\approx 0$ and $\rho^{\mathrm{eeg,llm},l}_s\approx 0$ but $\rho^{\mathrm{ac,llm},l}_s$ may be large;
in that regime,
\begin{equation}
\mathrm{TNC}^{\,l}_s \approx \frac{1}{3}\big[\rho^{\mathrm{ac,llm},l}_s\big]^2,
\label{eq:tnc_div3}
\end{equation}
i.e., the spurious pairwise alignment is down-weighted by a factor of $3$ in TNC.

This directly formalizes the main-text claim: TNC mitigates indirect/shared confounds that inflate only a subset of modality pairs,
because such effects do not produce uniformly high tri-modal coherence in $\mathbf{R}^{\,l}_s$ (Eq.~\ref{eq:tri_corr_matrix}).

\subsection{Why squaring helps (sign invariance) without changing the ``conservative'' logic}

Because RSA correlations can be negative under sign-reversing but geometry-preserving transformations (e.g., reversed rank order),
we square each Spearman RSA term in Eq.~\ref{eq:tnc}. Formally, for any pair $(m,n)$,
\begin{equation}
\big[\rho^{m,n}_s\big]^2 = \big[\rho^{-m,n}_s\big]^2,
\end{equation}
so TNC depends on the \emph{magnitude} of geometric agreement, not its arbitrary sign.
All results above (boundedness, near-1 implication, noise attenuation, and confound dilution) continue to hold because they are stated
in terms of squared correlations and variance-controlled magnitudes.

\subsection{Summary of the formal guarantees}

Combining Eqs.~\ref{eq:tnc_frobenius}--\ref{eq:tnc_div3} yields the key benefits claimed in the main text:

\begin{itemize}
\item \textbf{Tri-modal coherence:} TNC is a normalized off-diagonal energy of the tri-modal correlation matrix (Eq.~\ref{eq:tnc_frobenius}).
\item \textbf{High TNC requires all pairs high (in a precise sense):} if $\mathrm{TNC}^{\,l}_s$ is close to $1$, each pairwise RSA magnitude must be close to $1$ (Prop.~\ref{prop:near1}).
\item \textbf{Conservative under modality-specific noise:} if one modality is noisy, two of the three correlations are attenuated (Eqs.~\ref{eq:attenuation}--\ref{eq:tnc_expected_noise}), suppressing TNC even when a single pair remains high.
\item \textbf{Mitigates pair-specific confounds:} a confound shared by only two modalities inflates at most one term and often reduces the other two, yielding a diluted contribution to TNC (Prop.~\ref{prop:pair_confound_dilution}, Eq.~\ref{eq:tnc_div3}).
\end{itemize}

\section{Electrode-wise Enriched EEG Features: Definitions and Computation}
\label{app:enriched_eeg_features}

This appendix details the electrode-wise enriched EEG features used in the main text, including (i) what each feature captures,
(ii) the precise mathematical definition, and (iii) how it is computed on the token-aligned time grid.

\subsection{Setup.}
For each sentence $s$ and electrode $c\in\{1,\ldots,C\}$, we operate on the token-aligned EEG voltage sequence
$\{e_{s,t,c}\}_{t=1}^{T_a}$, where $e_{s,t,c}=\mathbf{U}^{\mathrm{eeg}}_s(t,c)$ and
$\mathbf{U}^{\mathrm{eeg}}_s\in\mathbb{R}^{T_a\times C}$.
All features below are computed \emph{after} temporal alignment so that each descriptor is defined on the same token-synchronous axis
used for EEG--model comparisons.

\subsection{Rationale: why these feature types?}
Our goal is to enrich each token-aligned EEG sample with complementary local descriptors that (i) remain well-defined on the
token-synchronous grid, (ii) are lightweight and robust to noise, and (iii) capture multiple aspects of short-term neural dynamics
that may be reflected in model representations~\citep{cohen2014analyzing,oppenheim2009dtsp}.
The selected feature set is intentionally minimal yet multi-view, consistent with the general motivation for combining diverse
time-series descriptors to cover complementary signal properties~\citep{fulcher2013hctsa}, and includes:

\textbf{(1) Instantaneous level} via $e_{s,t,c}$, preserving the raw potential needed to retain fine temporal detail~\citep{cohen2014analyzing}.

\textbf{(2) Local dynamics} via $\Delta e_{s,t,c}$ and $\Delta^2 e_{s,t,c}$, which summarize slope and curvature and thus emphasize
rapid changes (transient responses) that may be less visible in raw amplitude alone~\citep{oppenheim2009dtsp,cohen2014analyzing}.

\textbf{(3) Local context and stability} via $\mathrm{stats}_w(\cdot)$, which provides a short-range baseline and variability estimate
to mitigate token-level noise and capture slower local fluctuations on the aligned grid~\citep{oppenheim2009dtsp,cohen2014analyzing}.

\textbf{(4) Local power} via $\mathrm{RMS}_w(\cdot)$, summarizing sustained magnitude in a window and complementing signed voltages~\citep{oppenheim2009dtsp,cohen2014analyzing}.

\textbf{(5) Oscillatory content} via $\mathrm{BandPow}_w(\cdot)$, providing a coarse characterization of rhythmic activity that is not
fully captured by time-domain descriptors~\citep{allen1977stft,cohen2014analyzing}.

Together, these components form a compact descriptor that is expressive enough to capture level, change, local distribution,
power, and band-limited structure, while keeping the representation interpretable and compatible with electrode-wise similarity
analysis in the main text.

\subsection{Raw voltage (instantaneous amplitude).}
We include the raw voltage $e_{s,t,c}$ to retain the instantaneous potential level at token-aligned time $t$~\citep{cohen2014analyzing}.

\subsection{Discrete differences (local dynamics).}
Differences emphasize rapid changes, onset/offset-like transients, and local acceleration patterns that may be muted in raw amplitude.
The first difference captures local slope (fast rises/falls), while the second difference captures local curvature (changes of slope),
often reflecting short-lived events in the waveform~\citep{oppenheim2009dtsp,cohen2014analyzing}.

\paragraph{Definition and computation.}
We compute first- and second-order discrete differences as in the main text~\citep{oppenheim2009dtsp}:
\begin{equation}
\begin{aligned}
\Delta e_{s,t,c} &= e_{s,t,c} - e_{s,t-1,c},\\
\Delta^2 e_{s,t,c} &= \Delta e_{s,t,c} - \Delta e_{s,t-1,c}.
\end{aligned}
\end{equation}
At boundaries (e.g., $t=1$), $\Delta e_{s,t,c}$ and $\Delta^2 e_{s,t,c}$ require a convention.
In our implementation we apply the same boundary rule across all sentences and electrodes (e.g., padding/replication or masking).

\subsection{Windowed statistics $\mathrm{stats}_w(\cdot)$ (local baseline and variability).}
Sliding-window summaries stabilize the representation against token-level noise and summarize slower local context:
the windowed mean captures local baseline (including slow drift on the aligned grid),
the windowed standard deviation captures local variability,
and the windowed maximum captures local peak excursions (useful for transient bursts)~\citep{oppenheim2009dtsp,cohen2014analyzing}.

\paragraph{Window index set.}
Let $w$ be the window length (in token-aligned samples) and define the centered index set
\begin{equation}
\small
\begin{aligned}
\mathcal{I}_{t,w}
&=\Big\{\, i \in \{1,\ldots,T_a\} \;:\;\\
&\mathllap{t-\lfloor (w-1)/2\rfloor \le i \le t+\lceil (w-1)/2\rceil}\Big\}.
\end{aligned}
\end{equation}
where the intersection with $\{1,\ldots,T_a\}$ implements boundary truncation.
We denote the effective window size by $|\mathcal{I}_{t,w}|$.

\paragraph{Definition and computation.}
Over $\mathcal{I}_{t,w}$ we compute:
(i) windowed mean $\mu_{w,s,c}(t)$,
(ii) windowed standard deviation $\sigma_{w,s,c}(t)$,
and (iii) windowed maximum $m_{w,s,c}(t)$:
\begin{equation}
\small
\begin{aligned}
\mu_{w,s,c}(t)
&= \frac{1}{|\mathcal{I}_{t,w}|}\sum_{i\in\mathcal{I}_{t,w}} e_{s,i,c},\\[2pt]
\sigma_{w,s,c}(t)
&= \sqrt{\frac{1}{|\mathcal{I}_{t,w}|}\sum_{i\in\mathcal{I}_{t,w}}
\big(e_{s,i,c}-\mu_{w,s,c}(t)\big)^2},\\[2pt]
m_{w,s,c}(t)
&= \max_{i\in\mathcal{I}_{t,w}} e_{s,i,c}.
\end{aligned}
\end{equation}

We then define the windowed-statistics operator used in the main text as the stacked vector
\begin{equation}
\small
\begin{aligned}
\mathrm{stats}_w\!\big(e_{s,\cdot,c};t\big)
&=
\begin{bmatrix}
\mu_{w,s,c}(t)\\
\sigma_{w,s,c}(t)\\
m_{w,s,c}(t)
\end{bmatrix}
\in \mathbb{R}^{D_{\mathrm{stats}}},\\
D_{\mathrm{stats}} &= 3.
\end{aligned}
\end{equation}

\subsection{Windowed RMS $\mathrm{RMS}_w(\cdot)$ (local power).}
RMS summarizes local signal power and is insensitive to sign changes.
It complements raw amplitude by reflecting sustained magnitude within a short neighborhood, often tracking local ``energy'' even when oscillatory~\citep{oppenheim2009dtsp,cohen2014analyzing}.

\paragraph{Definition and computation.}
Using the same window $\mathcal{I}_{t,w}$, we compute
\begin{equation}
\small
\mathrm{RMS}_w\!\big(e_{s,\cdot,c};t\big)
=
\sqrt{\frac{1}{|\mathcal{I}_{t,w}|}\sum_{i\in\mathcal{I}_{t,w}} e_{s,i,c}^2}
\in\mathbb{R}.
\end{equation}

\subsection{Short-time band energies $\mathrm{BandPow}_w(\cdot)$ (oscillatory content).}
Band energies provide a coarse summary of oscillatory content, complementing time-domain descriptors.
They help distinguish cases where EEG--model similarity is driven by rhythmic structure (band-limited activity) rather than pointwise amplitude~\citep{cohen2014analyzing}.

\paragraph{Definition and computation.}
Let $\mathcal{B}$ denote a predefined set of frequency bands.
For each token time $t$, we take a short-time segment over a spectral window $\mathcal{I}_{t,w_{\mathrm{fft}}}$ on the token-aligned axis
(possibly using multiple $w_{\mathrm{fft}}$; resulting features are concatenated).
Let $x_{s,c,t}(n)$ denote the windowed segment (optionally multiplied by a taper $h(n)$) and let $X_{s,c,t}(k)$ be its DFT,
i.e., a standard short-time Fourier/periodogram-based spectral estimate~\citep{allen1977stft}.
For band $b\in\mathcal{B}$, the band power is computed by summing squared magnitudes over the corresponding frequency bins:
\begin{equation}
P_{b,s,c}(t) \;=\; \sum_{k \in \mathcal{K}(b)} \big|X_{s,c,t}(k)\big|^2,
\end{equation}
where $\mathcal{K}(b)$ maps band $b$ to DFT bins under the token-aligned sampling grid induced by the temporal alignment~\citep{allen1977stft}.

We define the band-power operator in the main text as the stacked vector
\begin{equation}
\small
\begin{aligned}
\mathrm{BandPow}_w\!\big(e_{s,\cdot,c};t\big)
&=
\begin{bmatrix}
P_{b_1,s,c}(t)\\
\vdots\\
P_{b_{|\mathcal{B}|},s,c}(t)
\end{bmatrix}
\in \mathbb{R}^{D_{\mathrm{band}}},\\
D_{\mathrm{band}} &= |\mathcal{B}|.
\end{aligned}
\end{equation}
with the understanding that if multiple spectral windows are used, their band-power vectors are concatenated.

\subsection{Per-token feature stacking and dimensionality.}
At each token time $t$ and electrode $c$, we stack instantaneous amplitude, local dynamics, windowed statistics,
local RMS power, and band-limited energy exactly as in the main text:
\begin{equation}
\mathbf{x}_{s,t,c}
=
\begin{bmatrix}
e_{s,t,c}\\
\Delta e_{s,t,c}\\
\Delta^2 e_{s,t,c}\\
\mathrm{stats}_w\!\big(e_{s,\cdot,c};t\big)\\
\mathrm{RMS}_w\!\big(e_{s,\cdot,c};t\big)\\
\mathrm{BandPow}_w\!\big(e_{s,\cdot,c};t\big)
\end{bmatrix}
\in \mathbb{R}^{D_r}.
\end{equation}
where the feature dimension is
\begin{equation}
\small
D_r \;=\; 1 \;+\; 1 \;+\; 1 \;+\; D_{\mathrm{stats}} \;+\; 1 \;+\; D_{\mathrm{band}},
\end{equation}
or larger if multiple spectral windows are concatenated.

\paragraph{Sequence construction (per electrode).}
For a fixed electrode $c$, we form the enriched feature sequence
$\mathbf{X}^{\mathrm{eeg}}_{s,c} = [\mathbf{x}_{s,1,c};\ldots;\mathbf{x}_{s,T_a,c}] \in \mathbb{R}^{T_a\times D_r}$.
Stacking all electrodes yields $\mathbf{X}^{\mathrm{eeg}}_{s} \in \mathbb{R}^{T_a\times D_r\times C}$.

\section{Layer-wise EEG--Model Similarity Trajectories}
\label{sec:appendix_layerwise}

\begin{figure*}[h]
    \centering
    \includegraphics[width=1\textwidth]{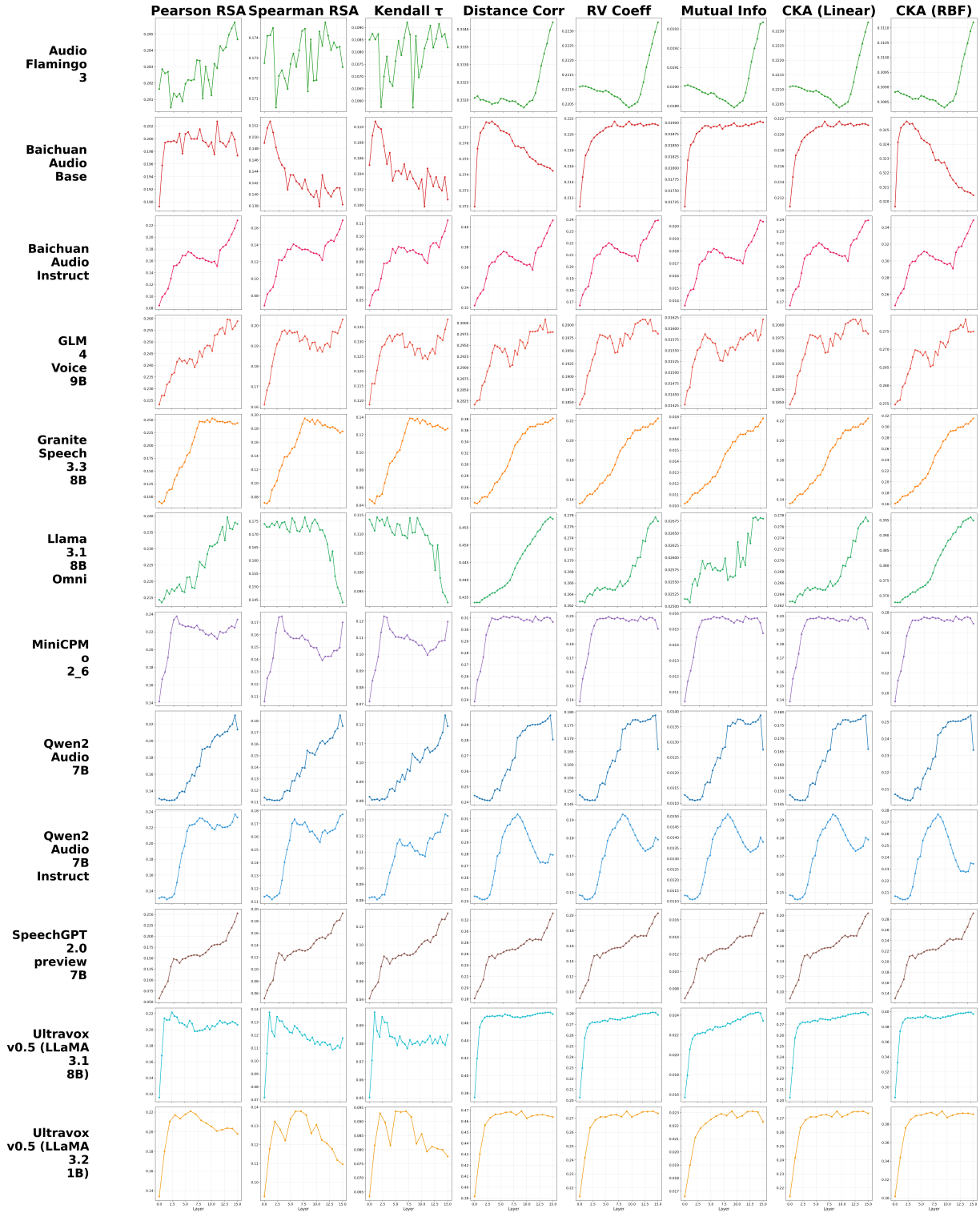}
    \caption{Layer-wise EEG--model similarity trajectories across 12 Audio LLMs.
    For each model (row) and metric (column), we compute similarity between EEG RDMs and layer-wise hidden-state RDMs after aligning model time steps to the EEG axis.
    Metrics include Pearson RSA $r$, Spearman RSA, Kendall $\tau$, distance correlation, RV coefficient, mutual information, and CKA (linear/RBF).}
    \label{fig:all_models_layerwise_grid}
\end{figure*}

\paragraph{How to read Fig.~\ref{fig:all_models_layerwise_grid}.}
For each Audio LLM, we extract hidden states at every transformer block (layer index is 0-based), temporally align the resulting sequence to the EEG time grid, and then compute representational dissimilarity matrices (RDMs) over aligned time steps. Cross-modal similarity is measured between vectorized RDMs.
The eight columns report complementary notions of alignment: Pearson RSA $r$ captures linear association; Spearman RSA and Kendall $\tau$ capture rank-order (geometry-preserving) agreement; distance correlation (dCor) and mutual information (MI) capture more general dependence beyond linearity; RV coefficient measures linear multivariate association between distance structures; and CKA (linear/RBF) measures representation similarity with linear and kernelized variants \citep{kriegeskorte2008rsa,kendall1945tiesties,kornblith2019cka,szekely2007dcor,robert1976rv,cover2006it}.
Because all curves are computed \emph{after} aligning model states to the EEG time axis, differences across layers primarily reflect representational transformations inside the model rather than temporal sampling mismatches.

\paragraph{Audio-Flamingo-3.}
AF3 exhibits comparatively \emph{layer-invariant} alignment: Pearson RSA and rank-based metrics fluctuate in a narrow band, while dependence-style metrics (dCor/CKA/MI/RV) show only a mild late-layer uplift.
This “flat” profile is consistent with AF3’s design where a strong audio encoder (AF-Whisper) and modality-bridging components feed a language backbone, yielding relatively stable time-wise geometry across blocks \citep{goel2025audioflamingo3}.
Empirically, Spearman RSA peaks near the upper-middle layers (around layer 21; Spearman RSA $\approx 0.175$) but changes little thereafter, suggesting that EEG-relevant geometry is already present once audio features have been injected into the backbone.

\paragraph{Baichuan-Audio-Base.}
The base checkpoint shows a marked \emph{early-peak then decay} pattern for rank-based geometry: Spearman RSA reaches its maximum extremely early (layer 2; Spearman RSA $\approx 0.153$) and then steadily declines across depth, with Kendall $\tau$ showing the same tendency.
Meanwhile, dependence measures (especially dCor/CKA-RBF) are strongest in early-to-mid layers and then drift downward, while RV/CKA-L and MI peak later (around layers 17 and 26; RV $\approx 0.222$, MI $\approx 0.019$), whereas Pearson RSA$r$ is comparatively stable (small net change from shallow to deep).
A plausible interpretation is that in the base model, early layers preserve perceptual/phonetic time geometry that resembles EEG tracking, while deeper layers become increasingly specialized for downstream generation/interaction objectives, reshaping rank structure in a way that is less EEG-like \citep{li2025baichuanaudio}.

\paragraph{Baichuan-Audio-Instruct.}
In sharp contrast to the base checkpoint, the instruction-tuned variant shows a \emph{progressive increase} across essentially all metrics, with peaks concentrated at late layers (final block: Spearman RSA $\approx 0.169$, dCor $\approx 0.407$, CKA-RBF $\approx 0.349$).
This indicates that instruction tuning can substantially reorganize depth-wise representations so that EEG-aligned geometry and dependence both \emph{accumulate} with depth, rather than being strongest only in early perceptual layers.
The base vs.\ instruct contrast suggests that alignment is not purely determined by the audio front-end, but is strongly modulated by how deeper layers are optimized for following instructions and producing grounded responses \citep{li2025baichuanaudio}.

\paragraph{Glm-4-Voice-9B.}
GLM-4-Voice displays one of the clearest \emph{late-layer consolidation} profiles: most metrics rise quickly in shallow layers and continue improving (with mild mid-layer ripples) until late depth, where Spearman RSA attains a high peak at the final layer (Spearman RSA $\approx 0.203$).
The consistency across Pearson RSA/RV/CKA and rank-based metrics suggests that GLM-4-Voice increasingly aligns both \emph{global dependence} and \emph{relative geometry} with EEG as depth increases.
This is compatible with an end-to-end speech-text architecture built on compact speech tokenization and joint modeling, where deeper blocks integrate longer-range linguistic structure \citep{zeng2024glm4voice}.

\paragraph{Granite-Speech-3.3-8B.}
Granite Speech shows strong monotonic improvements for dependence-oriented metrics (dCor/CKA/RV/MI), but rank-geometry metrics peak at \emph{intermediate} depth: Spearman RSA reaches its maximum around layer 16 (Spearman RSA $\approx 0.195$) and then slightly plateaus/decreases.
This separation suggests that later layers may still increase overall statistical dependence with EEG, while the \emph{ordering of pairwise dissimilarities} (geometry) becomes less stable, potentially reflecting specialization toward ASR/translation decoding objectives in top blocks.
Such an intermediate optimum is consistent with hierarchical processing views where mid-level representations carry substantial linguistic abstraction, while the final blocks are more task-head-biased \citep{saon2025granitespeech}.

\paragraph{Llama-3.1-8B-Omni.}
LLaMA-Omni exhibits a pronounced \emph{metric dissociation}: dependence measures rise strongly toward late depth (e.g., dCor peaks near layer 30 at $\approx 0.458$, CKA-RBF peaks $\approx 0.396$), yet Spearman RSA decreases after an intermediate peak (Spearman RSA max near layer 16 at $\approx 0.176$), while Kendall $\tau$ attains its absolute maximum very early (layer 3; Kendall $\tau \approx 0.114$) with a near-tie around layer 16 before declining toward late depth.
This indicates that deeper layers remain statistically coupled to EEG-like structure but increasingly \emph{re-rank} the time-wise geometry.
Given that Omni models are designed for speech interaction (often including response planning and/or speech generation pathways), the highest layers may encode generation- or turn-taking-related organization that is not present in passive listening EEG, thus reducing rank-order agreement despite higher overall dependence \citep{fang2024llamaomni}.

\paragraph{MiniCPM-o-2\_6.}
MiniCPM-o shows a characteristic \emph{fast early rise + non-monotonic middle} trajectory: Pearson RSA and Spearman RSA peak early (around layer 6; Pearson RSA $\approx 0.238$, Spearman RSA $\approx 0.175$), then dip and partially recover, while kernel/MI-style metrics reach maxima later (e.g., CKA-RBF peaks near layer 21).
This pattern is consistent with an omni setting where early blocks align speech features into a language-compatible manifold, whereas later blocks support broader multimodal/interactive behaviors that can reshape geometry (hence the mid-layer dip) while still increasing dependence measures \citep{openbmb2025minicpmo2626}.

\paragraph{Qwen2-Audio-7B.}
Qwen2-Audio displays a largely \emph{monotonic increase} across all metrics with a common peak at the penultimate layer (around layer 30: Pearson $\approx 0.231$, Spearman RSA $\approx 0.185$, MI $\approx 0.014$, CKA-RBF $\approx 0.254$), followed by a small drop at the final layer.
A penultimate-layer optimum is a known phenomenon in representation analyses, where the last block is most influenced by output-head constraints and task formatting, slightly distorting geometry while preserving strong semantic readiness one layer below the head \citep{chu2024qwen2audio}.

\paragraph{Qwen2-Audio-7B-Instruct.}
The instruction-tuned Qwen2-Audio variant remains strong in rank-geometry at late depth (Spearman RSA peaks at the final layer; $\approx 0.177$), but dependence metrics (dCor/CKA/MI) peak \emph{earlier} (around layer 17) and show a non-monotonic profile.
This again highlights that instruction tuning can redistribute where different “notions of similarity” concentrate: intermediate layers may maximize broad dependence, while the final blocks optimize instruction-following and output formatting in a way that preserves rank geometry but can reduce kernelized/global dependence relative to mid-depth.
Overall, the curve shapes suggest a functional split between “representation building” (mid depth) and “instruction-conditioned decision/readout” (top depth) \citep{chu2024qwen2audio}.

\paragraph{SpeechGPT-2.0-preview-7B.}
SpeechGPT shows a robust \emph{depth-wise accumulation} across essentially all metrics, with peaks at the final layer (Spearman RSA $\approx 0.193$, Pearson RSA $\approx 0.253$).
The consistency across rank and dependence metrics suggests that deeper layers jointly refine both geometry and global coupling with EEG.
This is compatible with SpeechGPT’s speech-text mixed modeling (e.g., codec/patchify style design and multi-stage training) that progressively integrates semantics and acoustics into a unified sequence representation \citep{openmoss2025speechgpt2previewpreviewpreview}.

\paragraph{Ultravox-v0.5 (LLaMA-3.1-8B).}
Ultravox exhibits a strong \emph{early-geometry / late-dependence} split: Spearman RSA peaks very early (layer 2; $\approx 0.138$) and then declines, while dependence measures (dCor/CKA/MI/RV) rise sharply and saturate at high values toward late depth (e.g., dCor $\approx 0.472$, CKA-RBF $\approx 0.399$).
This pattern is consistent with an adapter-based multimodal stack where an audio front-end and bridging modules create strong statistical coupling, but the frozen or instruction-centric LLM backbone reshapes rank ordering as depth increases, weakening Spearman RSA even when dependence grows \citep{fixie2025ultravox}.

\paragraph{Ultravox-v0.5 (LLaMA-3.2-1B).}
The smaller Ultravox model follows the same qualitative signature, with earlier saturation (fewer layers): Spearman RSA peaks around layer 7 ($\approx 0.138$), and dependence metrics reach their maxima by mid depth (dCor $\approx 0.469$, CKA-RBF $\approx 0.394$).
This suggests that scaling down compresses the depth at which audio-to-language coupling is established, but does not eliminate the rank-vs-dependence dissociation seen in the larger Ultravox \citep{fixie2025ultravox}.

\paragraph{Cross-model takeaway.}
Across models, we repeatedly observe that \emph{rank-based geometry} (Spearman RSA/Kendall) and \emph{dependence/CKA-style} metrics can peak at different depths, implying that “being correlated” and “preserving representational geometry” are not equivalent.
Architecturally, models optimized for end-to-end speech-text interaction and instruction following (e.g., GLM-4-Voice, SpeechGPT, Qwen2-Audio) tend to show late-layer improvements in Spearman RSA, whereas adapter-heavy or speech-to-speech oriented stacks (e.g., Ultravox, some Omni variants) more often show late-layer gains in dependence metrics but reduced rank-order agreement, plausibly reflecting a shift toward generation-specific organization in the top layers.

\begin{figure*}[h]
    \centering
    \includegraphics[width=1\textwidth]{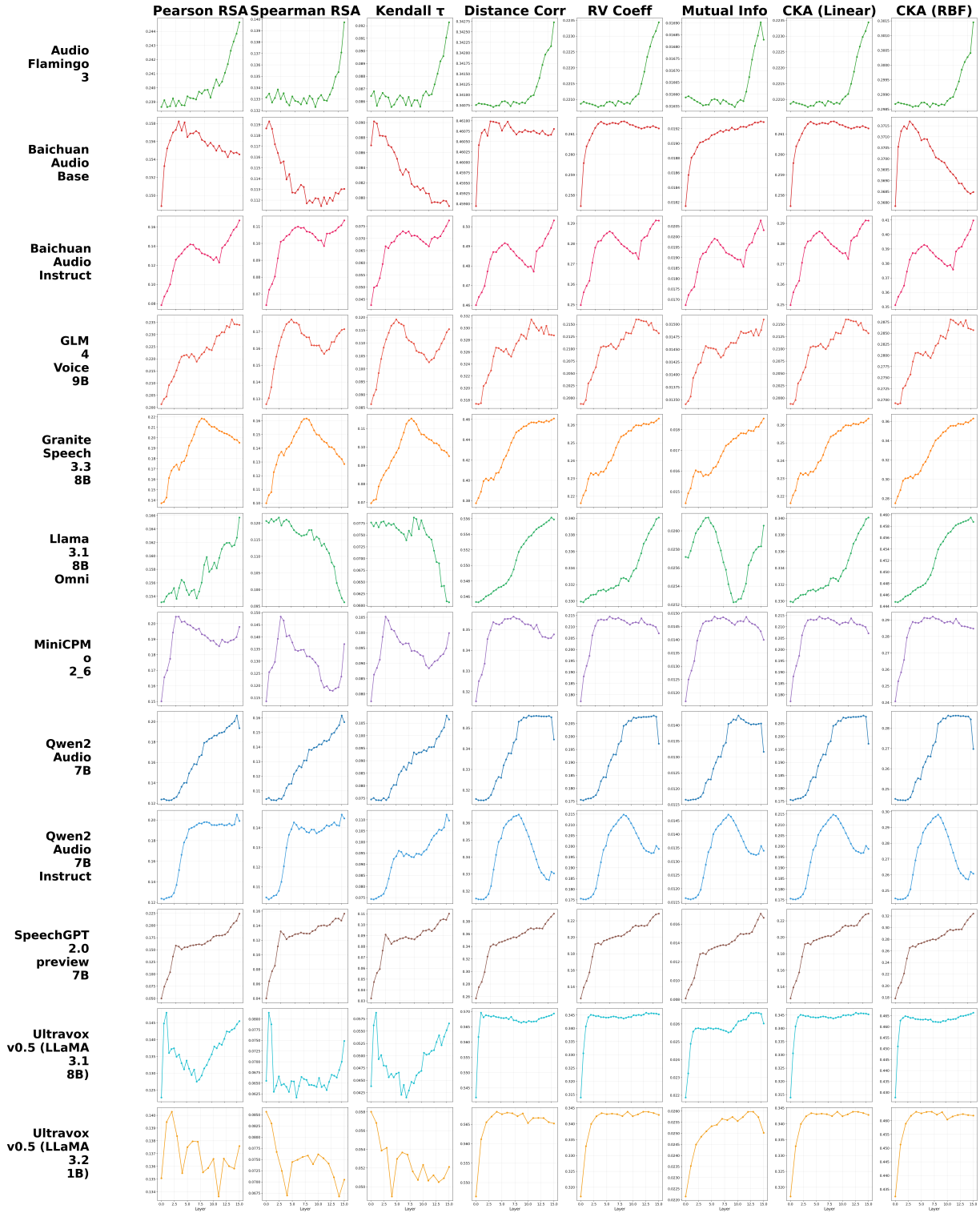}
    \caption{Layer-wise EEG--model similarity trajectories across 12 Audio LLMs.(cross-dataset replication on ds004408).}
    \label{fig:all_models_layerwise_grid2}
\end{figure*}
\paragraph{Cross-dataset replication on ds004408.}
To assess whether the layer-wise alignment trends in Fig.~\ref{fig:all_models_layerwise_grid} generalize beyond our in-house sentence-segmented dataset, we repeat the \emph{same} pipeline on the public continuous-speech EEG dataset ds004408 released via OpenNeuro \citep{openneurods004408,markiewicz2021openneuro}.
This dataset contains EEG from healthy adults listening to naturalistic audiobook speech (segments from \emph{The Old Man and the Sea}), recorded with a 128-channel BioSemi system at 512\,Hz \citep{bialas2025segmentation}.
As in the main analysis, we extract hidden states from each \textbf{LLM layer} of every Audio LLM, temporally align the model time axis to EEG, and then compute RDMs after alignment (i.e., the audio/model embeddings are first aligned to EEG time steps, then RDM/RSA and the complementary dependence metrics are computed).

\paragraph{Dataset-level comparison.}
Across models, the \emph{relative ordering} of “where alignment peaks” remains broadly stable (e.g., GLM-4-Voice and Granite-Speech still yield among the highest rank-based peaks), but the \emph{absolute magnitude} shifts: peak rank-based scores (Spearman RSA/Kendall) are systematically lower than in Fig.~\ref{fig:all_models_layerwise_grid}, whereas dependence-style scores (dCor/CKA/RV) are comparable or higher.
This cross-dataset dissociation is consistent with the idea that rank-based geometry agreement is more sensitive to dataset-specific noise, montage, and temporal variability, while dependence measures can remain high when the two modalities share strong global covariance structure even if pairwise similarity \emph{ordering} differs.

\paragraph{Audio-Flamingo-3.}
All metrics show a shallow early plateau followed by a clear late-layer rise, with peaks concentrated at the final layer (Spearman RSA peak $\approx 0.140$ at layer 27; dCor peak $\approx 0.343$ at layer 27), mirroring the “late-layer improvement” pattern in Fig.~\ref{fig:all_models_layerwise_grid} but at a reduced Spearman RSA scale.

\paragraph{Baichuan-Audio-Base.}
Rank-based metrics peak extremely early (Spearman RSA peak $\approx 0.119$ at layer 1) and then gradually decline, while dependence metrics stay high (dCor peak $\approx 0.461$ at layer 5; CKA-RBF peak $\approx 0.372$ at layer 5).
This reproduces the “early geometry match, later re-ordering” signature: deeper layers preserve strong shared variance with EEG but increasingly reshape the representational \emph{ordering} captured by Spearman RSA.

\paragraph{Baichuan-Audio-Instruct.}
Compared with the base model, instruction tuning shifts peaks to later layers (Spearman RSA peak $\approx 0.114$ at layer 27; dCor peak $\approx 0.503$ at layer 27), indicating that deeper LLM blocks contribute more strongly to EEG-aligned geometry on ds004408 than in the non-instruct counterpart.

\paragraph{GLM-4-Voice-9B.}
GLM retains the strongest rank-based peak among all models on ds004408 (Spearman RSA peak $\approx 0.177$ at layer 10), with a characteristic mid-layer maximum followed by a mild decrease.
In contrast, dependence metrics continue to increase toward deeper layers (e.g., CKA-RBF peaks later), suggesting a separation between (i) mid-layer representations that best match EEG’s \emph{rank-geometry} and (ii) late-layer representations that amplify global covariance.

\paragraph{Granite-Speech-3.3-8B.}
Granite shows a pronounced “rise-then-soft-fall” in rank-based scores (Spearman RSA peak $\approx 0.162$ at layer 16) alongside a strong monotonic increase in dependence measures (dCor peak $\approx 0.461$ at layer 31; CKA-RBF peak $\approx 0.363$ at layer 31).
This mirrors Fig.~\ref{fig:all_models_layerwise_grid}: intermediate layers are most EEG-consistent in ordering, while deeper layers increasingly concentrate shared variance.

\paragraph{Llama-3.1-8B-Omni.}
Spearman RSA peaks early (Spearman RSA peak $\approx 0.122$ at layer 5) and then steadily decreases toward the deepest layers, whereas dependence grows strongly (dCor peak $\approx 0.556$ at layer 30; CKA-RBF peak $\approx 0.460$ at layer 30).
This is a particularly clear example of “high dependence but low rank-consistency” in late layers.

\paragraph{MiniCPM-o-2\_6.}
MiniCPM exhibits a rapid early increase and a relatively stable plateau thereafter (Spearman RSA peak $\approx 0.148$ at layer 5), with dependence metrics saturating quickly.
Compared to models with strong late-layer growth, MiniCPM appears to concentrate EEG-relevant geometry in early-to-mid blocks and changes little afterward.

\paragraph{Qwen2-Audio-7B.}
Qwen2-Audio shows a largely monotonic improvement with depth (Spearman RSA peak $\approx 0.162$ at layer 30), consistent with Fig.~\ref{fig:all_models_layerwise_grid} and indicating that deeper layers continue to refine representations in a way that is increasingly EEG-consistent.

\paragraph{Qwen2-Audio-7B-Instruct.}
The instruct version preserves the general “late-layer best” trend (Spearman RSA peak $\approx 0.147$ at layer 30) but shows a more arched dependence trajectory (dCor/CKA rise then partially regress), suggesting that instruction tuning can redistribute where shared covariance is concentrated without eliminating the late-layer Spearman RSA advantage.

\paragraph{SpeechGPT-2.0-preview-7B.}
SpeechGPT remains strongly depth-progressive (Spearman RSA peak $\approx 0.156$ at layer 27; dCor peak $\approx 0.392$ at layer 27), aligning with the interpretation that later layers encode increasingly abstract (and brain-relevant) structure under naturalistic listening.

\paragraph{Ultravox-v0.5 (LLaMA-3.1-8B).}
Ultravox exhibits very low Spearman RSA overall with an early peak (Spearman RSA peak $\approx 0.081$ at layer 1) but extremely high dependence (dCor peak $\approx 0.570$ at layer 2; CKA-RBF peak $\approx 0.466$ at layer 31).
This “dependence $\gg$ Spearman RSA” regime suggests that, although model and EEG share strong global structure, the detailed pairwise similarity ordering implied by LLM-layer embeddings diverges substantially.

\paragraph{Ultravox-v0.5 (LLaMA-3.2-1B).}
The 1B variant shows the same qualitative dissociation (Spearman RSA peak $\approx 0.086$ at layer 0; dCor peak $\approx 0.568$ at layer 4), indicating that this pattern is not purely a scale effect but likely reflects how this model family organizes post-audio representations across depth.

\section{Supplementary Heatmaps and Paired RDM Visualizations}
\label{sec:Supplementary Heatmaps and Paired RDM Visualizations}
\begin{figure*}[h]
    \centering
    \includegraphics[width=1\textwidth]{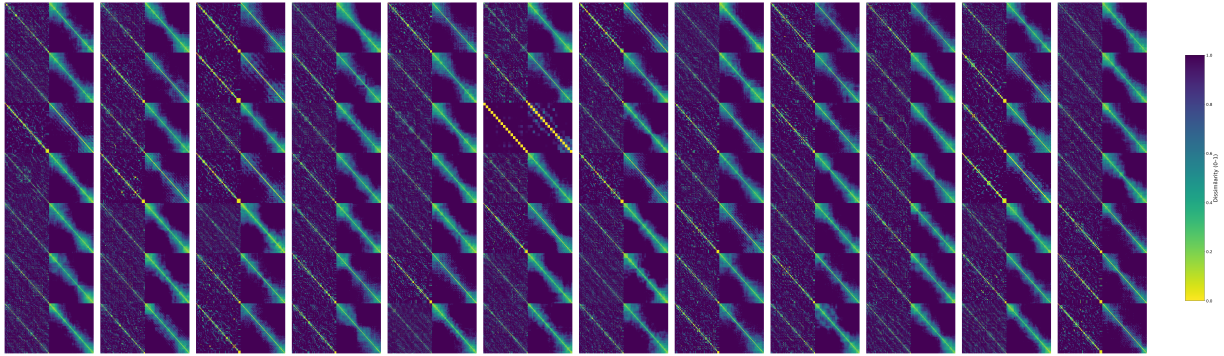}
  \caption{Paired representational dissimilarity matrices for audio and EEG across all 84 sentences.}
  \label{fig:app_rdm_pairs}
\end{figure*}

\begin{figure*}[h]
    \centering
    \includegraphics[width=1\textwidth]{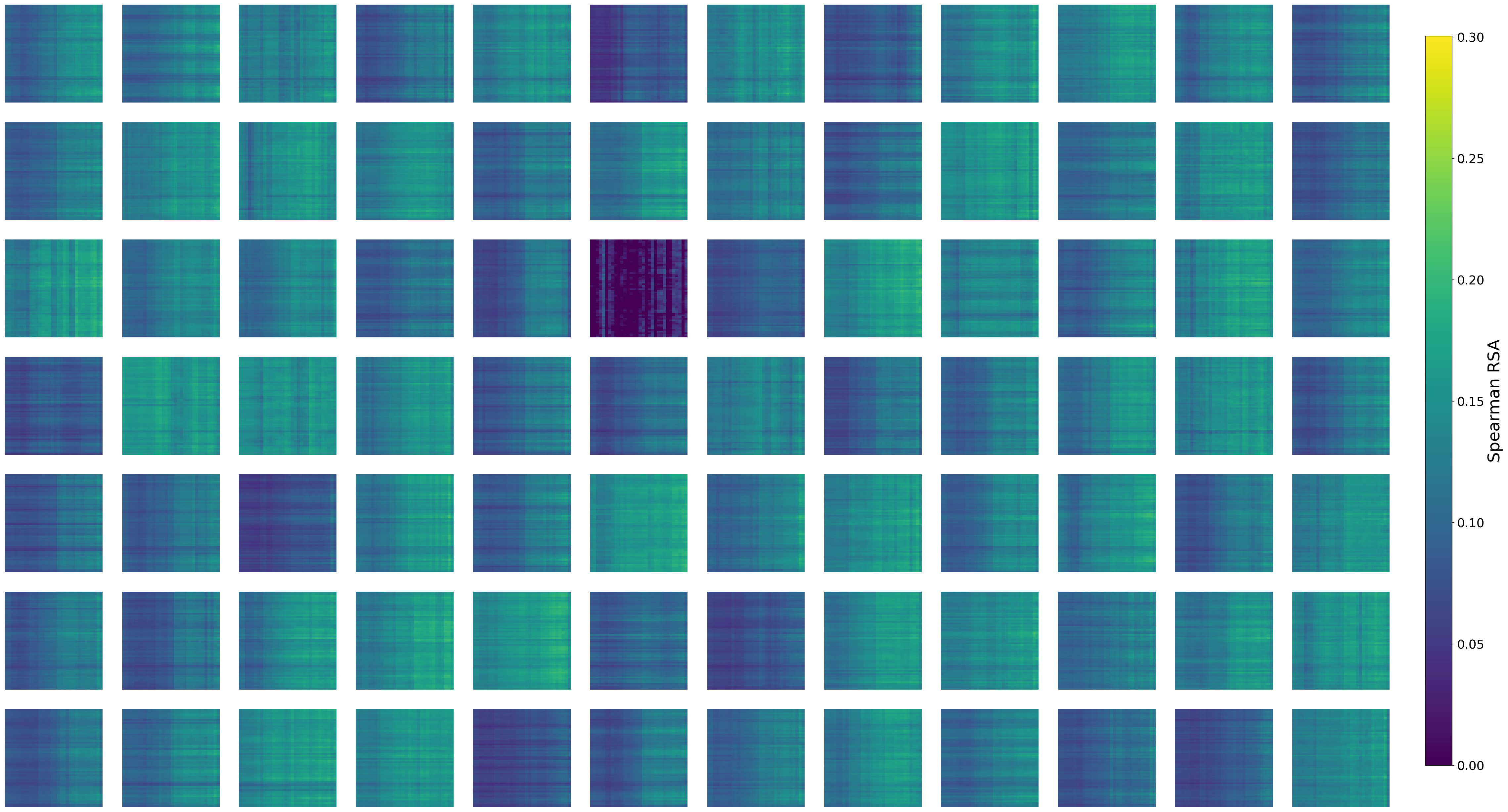}
  \caption{The heatmap results between electrodes and audio--model LLM layers across all 84 sentences.}
  \label{fig:app_spearman_heatmaps}
\end{figure*}

\paragraph{Correspondence to main-text figures.}
Figures~\ref{fig:app_spearman_heatmaps} and~\ref{fig:app_rdm_pairs} provide the complete set of sentence-level visualizations for Dataset~1, corresponding to the representative examples discussed in the main text (Fig.~\ref{fig:heatmap} and Fig.~\ref{fig:rdm_pair}, respectively). Panels are ordered left-to-right and top-to-bottom following the original sequence of the 84 sentence stimuli. In Fig.~\ref{fig:app_rdm_pairs}, each panel shows paired correlation-distance RDMs computed on the token-aligned time grid (audio--model RDM on the left, EEG RDM on the right). In Fig.~\ref{fig:app_spearman_heatmaps}, each panel shows the electrode-by-layer Spearman RSA heatmap between EEG features and Audio LLM hidden states, plotted with a shared color scale to enable direct comparison across sentences.

\end{document}